\documentclass[runningheads]{llncs}
\usepackage{graphicx}
\usepackage{cite}
\usepackage{url}
\usepackage{marvosym}
\usepackage{cite}
\usepackage{amsmath,amssymb,amsfonts}
\usepackage{algorithmic}
\usepackage{graphicx}
\usepackage{textcomp}
\usepackage{graphicx}
\usepackage[numbers]{natbib}
\usepackage{amssymb}
\usepackage{graphicx}
\usepackage{textcomp}
\usepackage{caption}
\usepackage{color}
\usepackage[dvipsnames]{xcolor}
\usepackage{subfigure}
\usepackage{multirow}
\usepackage{rotating}
\usepackage{diagbox}
\usepackage{stfloats}
\usepackage{float}
\usepackage{bm}
\usepackage{booktabs}
\usepackage{float}
\usepackage{multirow}
\begin{document}

\title{A Survey of Semen Quality Evaluation in Microscopic Videos Using Computer Assisted Sperm Analysis}
\authorrunning{W. Zhao et al.}
\titlerunning{A Survey of Semen Quality Evaluation}        

\author{Wenwei Zhao\inst{1}\and
Pingli Ma\inst{1}\and
Chen Li\inst{1 \textsuperscript{\Letter}}\and
Xiaoning Bu\inst{2}\and
Shuojia Zou\inst{1}\and
Tao Jiang\inst{3 \textsuperscript{\Letter}}\and
Marcin Grzegorzek\inst{4}}

\institute{Microscopic Image and Medical Image Analysis Group, MBIE College, 
Northeastern University, Shenyang, 110169, China \\
\and 
University of Washington (Seattle Campus), Seattle, US \\
\and
School of Control Engineering, Chengdu University of Information Technology, 610225, Chengdu, PR China\\
\and
Institute of Medical Informatics, University of Luebeck, Germany}

\date{Received: date / Accepted: date}

\maketitle

\begin{abstract}
The \emph{Computer Assisted Sperm Analysis} (CASA) plays a crucial role in male reproductive health diagnosis and Infertility treatment. With the development of the computer industry in recent years, a great of accurate algorithms are proposed. With the assistance of those novel algorithms, it is possible for CASA  to achieve a faster and higher quality result. Since image processing is the technical basis of CASA, including pre-processing,feature extraction, target detection and tracking, these methods are important technical steps in dealing with CASA. The various works related to Computer Assisted Sperm Analysis methods in the last 30 years (since 1988) are comprehensively introduced and analysed in this survey. To facilitate understanding, the methods involved are analysed in the sequence of general steps in sperm analysis. In other words, the methods related to sperm detection (localization) are first analysed, and then the methods of sperm tracking are analysed. Beside this, we analyse and prospect the present situation and future of CASA. According to our work, the feasible for applying in sperm microscopic video of methods mentioned in this review  is explained. Moreover, existing challenges of object detection and tracking in microscope video are potential to be solved inspired by this survey.
\end{abstract}

\footnotetext[1]{
First author: Wenwei Zhao, E-mail: zhaowenwein@163.com;\\
Co-first author: Pingli Ma, E-mail: 1045086571@qq.com;\\
Corresponding author: Chen Li, E-mail: lichen201096@hotmail.com;\\
Co-corresponding author: Tao Jang, E-mail: jiang@cuit.edu.cn.\\
}

\section{Introduction}
\label{s:int} 
\subsection{Purposes of Sperm Analysis}
\label{ss:int:Purposes}
It is of great significance for human beings to have the next generation. However, due to the unhealthy lifestyle of many male, their sperm quality is usually no good enough for fertilizing their wife. According to the research of \cite{Practice-2012-DEIM}, one in six couples fail to have a baby. $30\%$ of them are caused by male infertility. Therefore, the sperm analysis based on\textit{Computer-Aided Sperm Analysis} (CASA) is meaningful for human beings. Besides this, for the agricultural trade, a good fertility rate for livestock would be a huge economic gain, which means employing CASA in this field is feasible and significant.

At present, most infertile couples realize their desire to have the next generation through \textit{In-Vitro} Fertilization (IVF). Especially, \textit{Intracytoplasmic Sperm Injection} (ICSI) is an effective way to solve the problem of infertility. However, due to the problem that the embryologist and technicians cannot select a sound sperm \cite{Leung-2010-ASII}, IVF always has a high failure rate of about 60$-$70$\%$.

According to research of the World Health Organization (WHO) \cite{WHO-1999-WLME}, quality of sperm can be measured by their movement types. Some specific criteria designed by WHO in 2010 are employed for sperm motility classification, seen in Tab.\ref{tab:tableWHO}. The classification of sperm motility is mainly determined by the speed and direction of sperm motility. 
\begin{table}[htbp!]
\centering
\caption{Human sperm motility classification by World Health Organization 
(WHO)~\cite{WHO-1999-WLME}.}
\label{tab:tableWHO}
\setlength{\tabcolsep}{4pt}
\resizebox{11cm}{!}{
\begin{tabular}{@{}cccc@{}}

\toprule
No. & Grade & Name                       & Movements($\mu$m/s)              \\ \midrule
1   & A     & Rapid progressive motility & above 25                               \\
2   & B     & Slow progressive motility  & 5 $-$ 25 \\
3   & C     & Non-progressive motility   & 0 $-$ 5               \\
4   & D     & Immotility                 & 0 \\ \bottomrule
\end{tabular}}
\end{table}

Traditional method of sperm quality evalution relies on manually counting through a microscope by experts. To achieve a detection result with more objective, some well-designed standards to determining which category the counted sperm belong to are necessary \cite{Menkveld-2010-CSLN}. In spite of this, the traditional method has some disadvantages, such as time consuming, highly subjective and low accuracy \cite{Tmlinson-2013-SQIR}. Considering these disadvantages of traditional method and the importance of sperm quality assessment, it is particularly meaningful to employ CASA in this field.

In~\cite{holt-1994-RCSA}, a review about ``assessing a single (donor) sample by the use 
of five CASA systems'' is introduced, where it reveals that errors arising through 
variations in operator expertise and sample handling had considerably more influence upon 
the data than the more subtle differences between computer systems. Similarly, the work 
of~\cite{verstegen-2002-CASA} compares the Sperm Quality Analyzer IIC variables with the 
CASA estimates. According to these two reviews, CASA technology has better objectivity 
and accuracy than the traditional manual testing method. The introduction of CASA techniques 
into clinical laboratories should therefore help to improve laboratory standardization and 
the implementation of quality control procedures. With the development of CASA technology, 
CASA estimates, which will be more mature and provide much help for quantitative diagnosis, 
CASA system will become one of the important means to diagnose 
infertility~\cite{larsen-2000-CSAP}.

\subsection{Motivation of This Survey}
\label{ss:int:Motivation}
\begin{figure}[htbp!]
\centering
\includegraphics[width=0.78\linewidth]{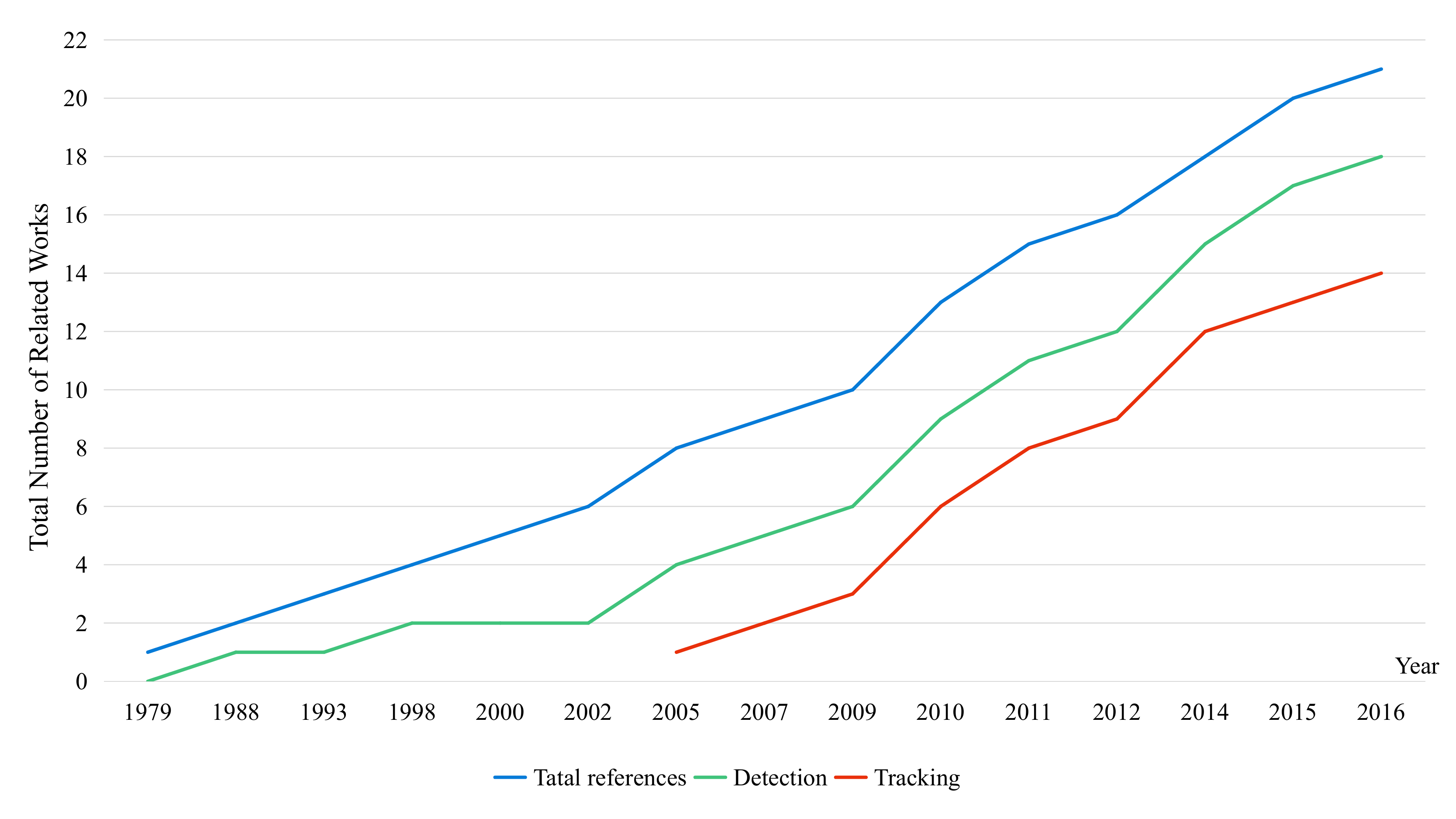}
\caption{Development trend of CASA over recent decades.}
\label{fig:Trend}
\end{figure}

We have first consult relevant works of literature and reviews about sperm quality analysis, 
and we find that most of them have the problems of containning too few or too old works of 
literature, and do not summarize or discuss the technology of using CASA. In the work 
of~\cite{Abbiramy-2010-SDCT}, it only summarizes nine papers from 2003 to 2010, and only 
compares simple applications of image processing, but it does not discuss the technical 
difficulties of CASA. In the survey of~\cite{berezansky-2007-STHS}, only six works of 
literature about sperm video segmentation and detection are mainly discussed, and no in-depth 
analysis of the detailed methods is given. To this end, we organize this survey paper to 
summarize about 20 related works on CASA from 1988 to 2017, covering an extensive time span 
and focusing on the techniques used in the CASA field. As shown in Fig.\ref{fig:Trend}, 
with the increase of infertility rate and the decline of sperm quality in recent years, 
people's attention and expectations for computer-assisted sperm quality analysis also show 
an increasing trend. 

\subsection{General Techniques in CASA Field}
\label{ss:int:Chart}

The main steps for CASA include: 
Specimen preparation, sperm segmentation and detection (localization), multi-sperm tracking and data association, and calculation of motility parameters. Fig.\ref{fig:FC} illustrates the general processing flow of CASA. Among this flow chart, commonly used algorithms in sperm detection and tracking are respectively exhibited. 

As shown in Fig.~\ref{fig:FC}:

\begin{figure*}[htbp!]
\centering
\includegraphics[width=0.78\linewidth]{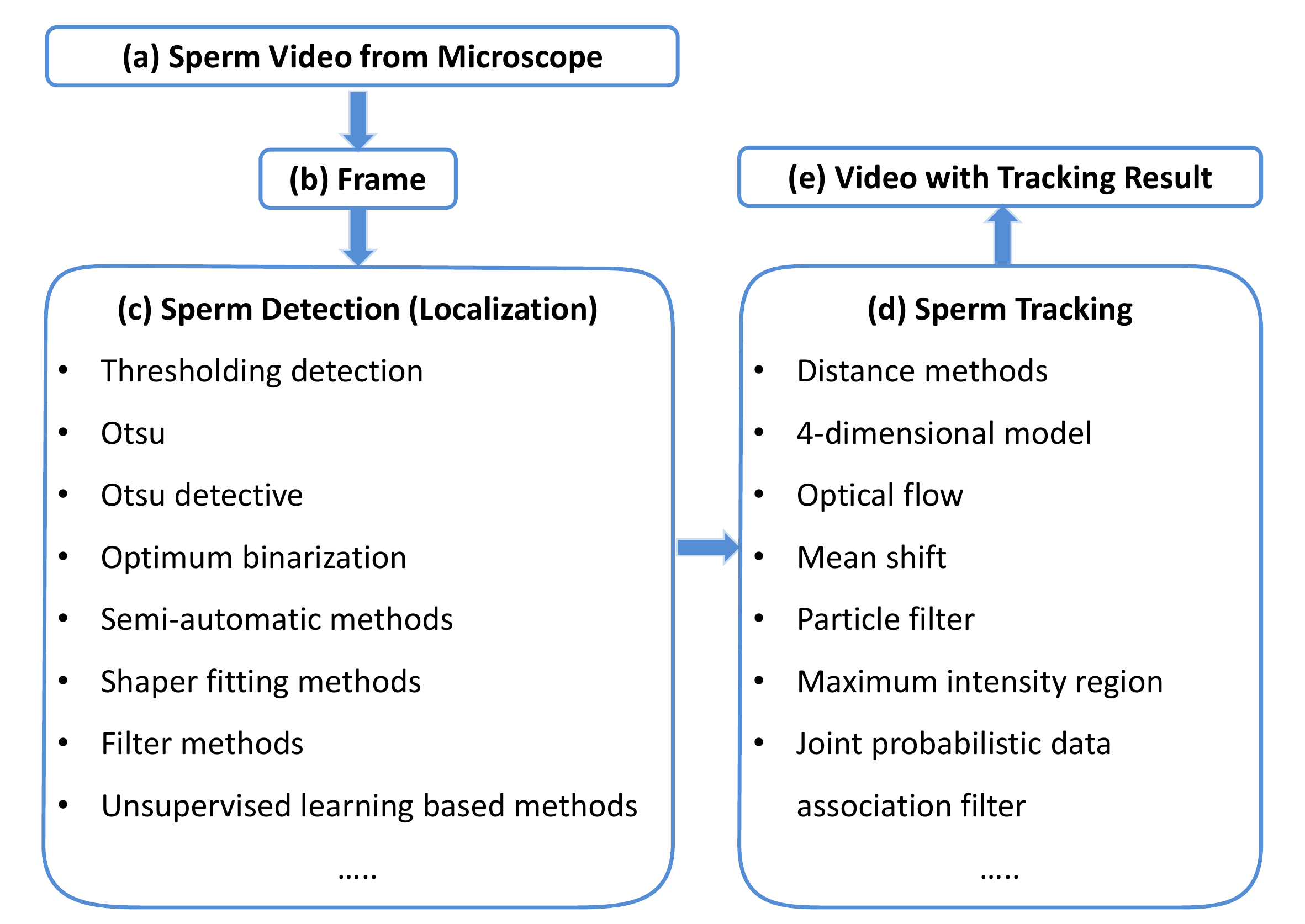}
\caption{A flowchart for general techniques in CASA field.}
\label{fig:FC}
\end{figure*} 

\begin{itemize}
\item[$\cdot$] The first step of CASA is video acquisition. Most of the research teams' 
microscopic videos are provided by hospitals and research institutes. There are also a 
small number of groups whose experimental data are provided by volunteers.

\item[$\cdot$] CASA system is an image-based research work. Therefore, the second step 
of the experiment is to frame the microscopic video so as to facilitate the subsequent 
detection and tracking of the experiment.

\item[$\cdot$] The third step is to detect or locate sperm. The main methods of detection 
and location include: Thresholding based detection, Otsu methods, optimum binarization 
methods, semi-automatic methods and so on. Details will be covered in Section~\ref{s:Detection}.

\item[$\cdot$] The fourth step of CASA is to track the sperm after detection. There are 
many common tracking algorithms applied to this step, such as mean shift, optical flow 
algorithm, particle filter tracking algorithm, and so on. There are also tracking algorithms 
developed by research teams themselves. These tracking algorithms are described in detail 
in Section~\ref{s:Tracking}.

\item[$\cdot$] The final step in CASA is to restore the images of the tracking results to video.
\end{itemize}

\subsection{Structure of This Survey}
\label{ss:int:Structure}
This paper is organized as follows: 
Section~\ref{s:Related}, some related knowledge about CASA is introduced. 
In Section~\ref{s:Detection}, we discuss sperm segmentation and detection (localization) 
techniques, and show representative images that demonstrate their operation. 
In Section~\ref{s:Tracking}, we summarize sperm tracking algorithms. 
Section~\ref{s:future} analyzes and prospects the present situation and future of CASA. 
In Section~\ref{s:Conclusion}, we close this survey with a brief conclusion.

\section{Related Knowledge about CASA}
\label{s:Related}

\subsection{Overview of CASA System Components}
\label{ss:Related:components}

In~\cite{boyers-1989-ASA} (section 3), the objectives of the CASA system in each stage 
are described in detail: (1) A successive image of the sperm suspension is projected onto 
the imaging detector; (2) sperm cells within each frame are detected based on pixel 
intensity and light scattering; and (3) professional software is used for obtaining necessary information and outputting the final result. To achieve the above objectives, the CASA 
system generally consists of four parts: Mechanical stage, illumination and optical systems, 
image acquisition (capture) and software output measures~\cite{amann-2014-CSAC}.

\textbf{Mechanical stage:} the optical microscope is moved to a predetermined position 
in the plane and autofocus is realized longitudinally.

\textbf{Illumination and optical systems:} CASA generally uses broadband lighting within 
the visible spectrum (390-700 nm). Pulse lighting is also typical in some systems, which 
cannot only meet the needs of the solid-state camera sensor shutter high-speed operation, 
but also can reduce the damage of sperm exposed to ultraviolet radiation. In addition, 
fluorescent lighting is also used in some devices to detect sperm stained with fluorescent 
pigments, to observe sperm morphology and whether there is membrane damage.

\textbf{Image acquisition (capture):} The acquisition of microscopic sperm video is 
accomplished by solid-state camera sensors (CCD or CMOS) in the hardware system. The 
smoothness of the image after framing is determined by the running speed of the camera 
shutter, which in turn affects the distance and path of sperm moving between successive 
frames. These parameters directly impact the output of the system and determine the accuracy 
of the CASA system diagnosis. The faster the shutter runs, the higher the video frames, 
and the closer the sperm motion-path is to the real situation, thus making the diagnosis 
more accurate.

All the existing CASA systems employ professional software \cite{amann-2014-CSAC}. There are mang open-source, such as Wilson-Leedy and Ingermann \cite{wilson-2007-DOAN}. This procedure mainly includes sperm cell detection, 
tracking, and sperm motility parameter calculation. Moreover the system outputs the results, 
which serve as the diagnostic criteria for sperm detection.

\subsection{Introduction to Abnormal Sperm}
\label{ss:Related:abnormal}
\begin{figure}[htbp!]
\centering
\includegraphics[width=0.75\linewidth]{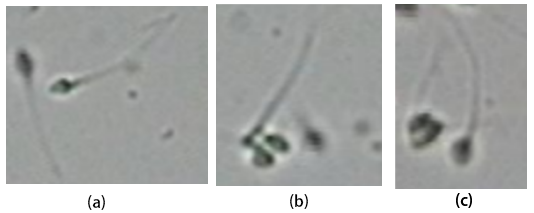}
\caption{ Example of three kinds of sperms.}
\label{fig:AS}
\end{figure}

Sperm health is mainly determined by two aspects: (1) Sperm motility parameters. 
For example, the moving speed mentioned in Tab.~\ref{tab:tableWHO}; (2) Sperm morphology. 
Sperm morphology has a decisive influence on sperm motility and movement speed. 
Morphologically, the majority of abnormal sperm can be roughly divided into two types: 
Double-headed or multi-headed sperm and malformed sperm. The comparison between normal 
sperm and abnormal sperm is shown in Fig.~\ref{fig:AS}, 
(a) normal sperm, 
(b) double-headed sperm, 
and (c) malformed sperm.

\subsection{Characteristics of Sperm in Microscopic Videos }
\label{ss:Related:characteristics}
\begin{figure}[htbp!]
\centering
\includegraphics[width=0.6\linewidth]{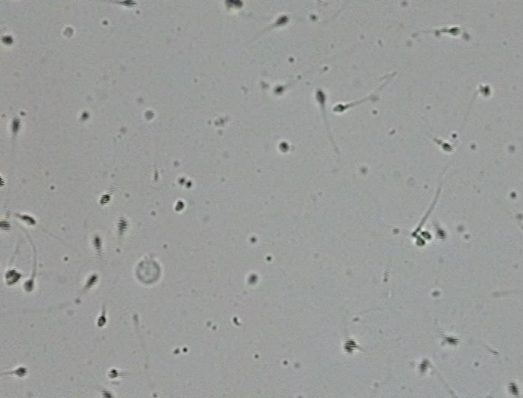}
\caption{ An example of a microscopic video (a frame) of sperm.}
\label{fig:SI}
\end{figure}

The microscopic video processing of sperm is a complicated and challenging process. 
In the microscopic video of sperm, sperm cells are usually colorless or transparent, 
so the color and texture information is basically invalid in target detection. In addition, 
each sperm cell has a highly similar morphology, which dramatically reduces the reliability 
of tracking algorithms in extracting information from morphology. At the same time, the 
video background is doped with many protein blocks or other impurities, which puts forward 
higher requirements for the early processing of video images. Less color information, similar 
morphology, and more impurities are the characteristics of sperm microscopic video. 
An example of a microscopic video (a frame) of sperm is shown in Fig.\ref{fig:SI}.

\section{Sperm Detection (Localization)}
\label{s:Detection} 
Sperm detection is the first step of CASA, and its accuracy is of great influence on the feasibility of the following tracking algorithm. The main purpose of sperm detection 
is to identify and locate sperm cells. For sperm microscopic videos, there are dozens or 
even hundreds of tiny and fast swimming sperm cells in each video frame, mixed with protein 
blocks and other impurities. Therefore, high-quality detection and recognition method is 
the key to making the tracking algorithm play its role. Efficient detection is like the 
foundation of a tall building. A solid foundation makes the project go smoothly. Otherwise, 
everything will be in vain.

To the best of our knowledge, the traditional detection and localization methods are the most 
frequently used, including thresholding-based, edge-based, 
and region-based segmentation techniques. Based on the selection of thresholding values to turn 
a grayscale into a binary image~\cite{Bhargavi-2014-STBS}, thresholding-based techniques is the 
simplest methods in segmentation. For example, Otsu thresholding, iterative thresholding, global 
thresholding, local thresholding and multi-thresholding approaches are regularly used methods. 
The key idea of edge based algorithms are converting original image into edge binary image based on abrupt changes in gray values \cite{Senthilkumaran-2009-EDTI}, such as Canny edge detector and active contour. Therefore, the selection of different edge detection methods 
also has a crucial impact on the experiment.

There are specific segmentation techniques that can only be used in specific technical 
situations~\cite{Pal-1993-RIST}. Below, we will summarize the detection and localization methods 
of sperm microscopic video analysis.

\subsection{Thresholding-based Segmentation Methods}
\label{ss:Detection:threshold}
In \cite{Kraemer-1998-FIHS}, the key idea of sperm detection is analysis the the influence of different threshold on sperm movement parameters. In this experiment, 650 levels of all 2000 gray levels are tested and analysed.
In \cite{Mack-1988-QSPM}, a thresholding-based segmentation technique is applied to process the sperm micro-video images. Especially, the judgement of which object segmented is identified as sperm head is determined by its size and gray value. After that, the center position of sperm head is obtained by analysing all the coordinates in this region.
\par
In \cite{nafisi-2005-SIUE}, an automatic sperm detection algorithm based on two-step threshold method is proposed. Firstly, a whole semen image is classified into foreground region and background region by employing a Otsu algorithm in global image histogram. After that, for obtaining detection result with higher precision, a Otsu algorithm is applied on a small square box that contains a particle, instead of a whole image.
\par
In \cite{leung-2010-DTLC}, sperm head region of interest (SHROI) is producted from original sample. First, center of SHROI is locadetd at where mouse points on. After that, an Otsu adaptive threshold algorithm \cite{otsu-1979-TSMF} is employed for SHROI binarization. Third, to improve the performance of final detection, a morphological close operation are applied as the post-processing. Finally, the processed SHROI is converting into binary image, where object is setted to one and background is setted to zero. After all the operations are done, centroid of sperm head is located by analysing outline of the sperm head.
\par

\begin{figure}[htbp!]
\centering
\includegraphics[width=0.65\linewidth]{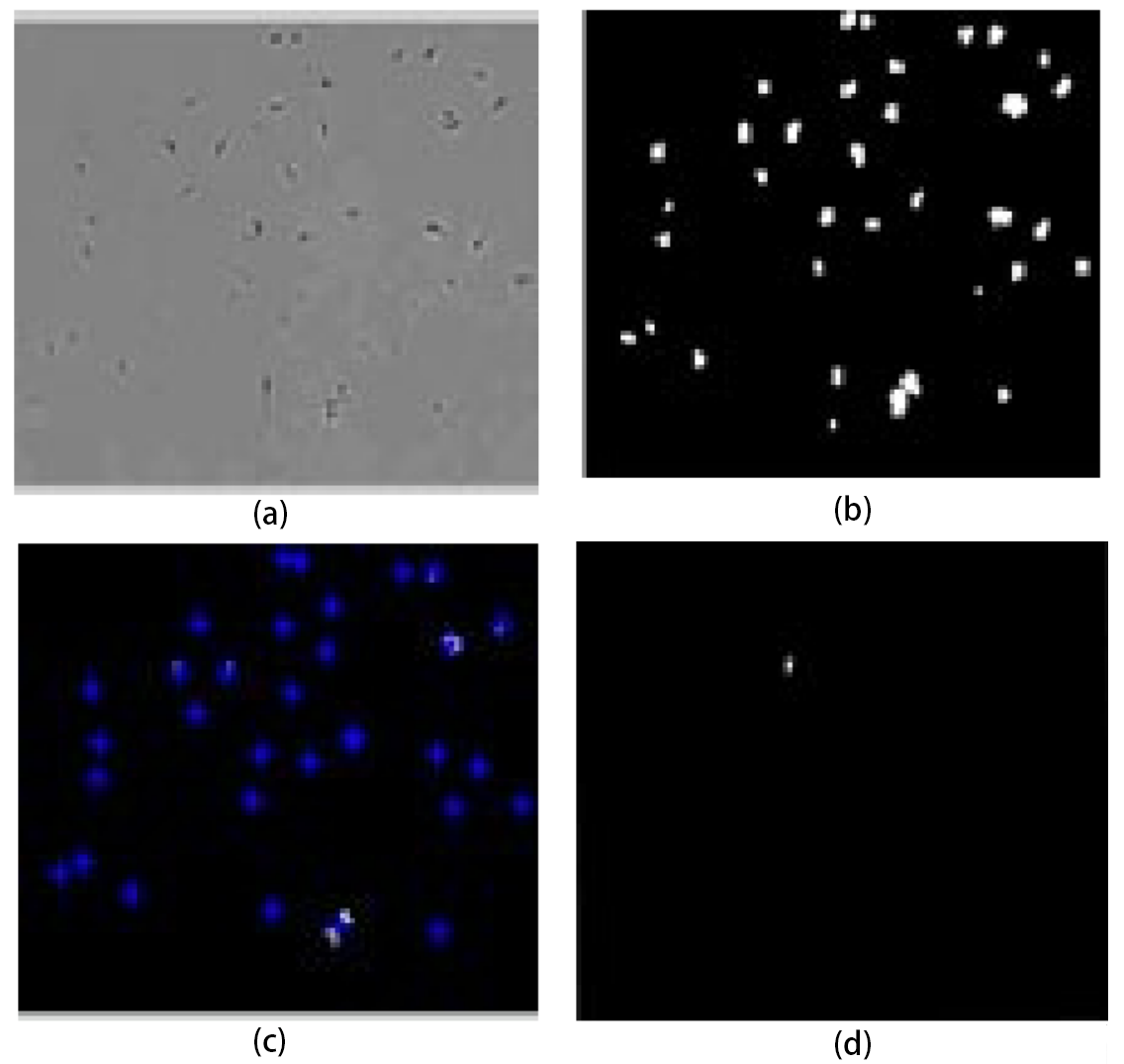}
\caption{The detection results of proposed method in \cite{Abbiramy-2010-SDCT}. The figure corresponds to Fig.4 in the original paper.}
\label{fig:SD}
\end{figure}

In \cite{Abbiramy-2010-SDCT}, the main processing of sperm detect consists of four steps. Firstly, the input image is binarized by employing Sobel operator. Then an operation of morphological dilation is applied for eliminating line-shape noise. Third, all the objects related to image boundary are deleted. At last, operation of morphological erosion is performed twice to smooth the detected object. An example of this work is shown in Fig.\ref{fig:SD}.
\par
In \cite{tomlinson-2010-VNCS}, a threshold segmentation algorithm and some morphological operations are combined to detect sperm. A threshold segmentation algorithm is first applied to distinguish between the sperm region and the background region. In addition, sperm regions with possible single sperm are specially detected in this step. After that, morphological operation is performed twice. One is morphological erosion to eliminate noise and other debris. The other is morphological dilation to smooth sperm regions detected and filling small holes. At last, all the candidate regions are screened based their area informations. Only those candidate regions which area are similar in size to the sperm head area are retained. In addition, for achieving higher accuracy, manual operation is applied to remove incorrectly screened regions.
\par
In \cite{lu-2011-robotic}, to get the detection target, a SHROI is first generated from initial image. The position of SHROI center  depends on where the mouse points to. After obtaining a SHROI, Otsu adaptive threshold method is employed for image binarization to highlight sperms in SHROI. At last, edges of detected sperms in SHROI are obtained and analysed.
\par
In \cite{liu-2012-QALB}, two consecutive frames are subtracted to obtain the boundary of sperm for target detection. After that, to make the boundary of sperm more clear, a threshold method is employed for eliminating noise and image binarization.
\par
\begin{figure}[htbp!]
\centering
\includegraphics[width=0.75\linewidth]{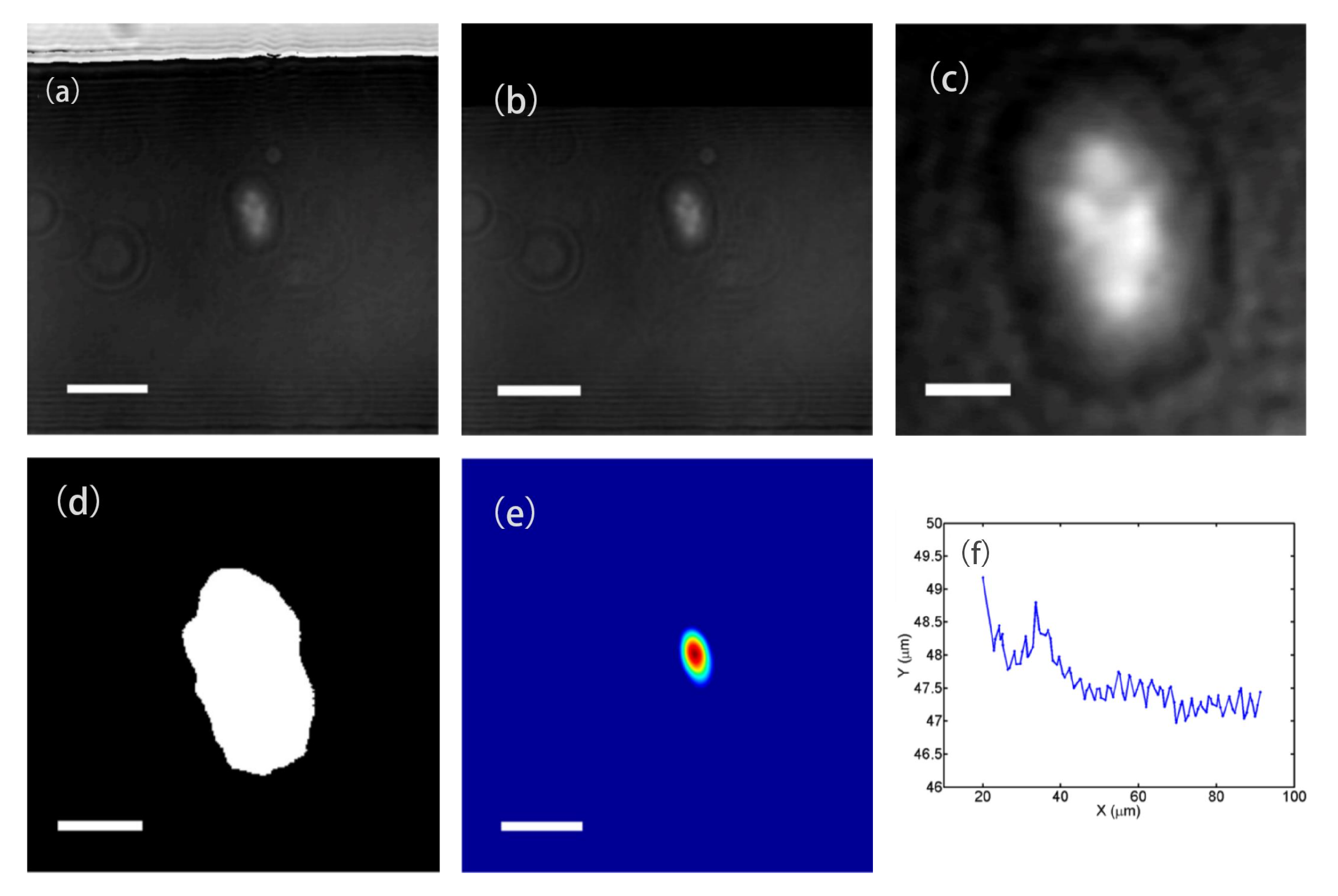}
\caption{The processing result for each step of proposed method in \cite{di-2014-4dTC}. The figure corresponds to Fig.2 in the original paper.}
\label{fig:4d}
\end{figure}

In \cite{di-2014-4dTC}, sperm detection part mainly consists of four steps. First, correspondent compensated phase map is obtained from original image. Second, Otsu threshold algorithm is applied for generating binary image. After automatically selecting regions of interest, the position of sperm is located based on selected function. Result of proposed detection method is shown in Fig.\ref{fig:4d}.

\par
\begin{figure}[htbp!]
\centering
\includegraphics[width=0.75\linewidth]{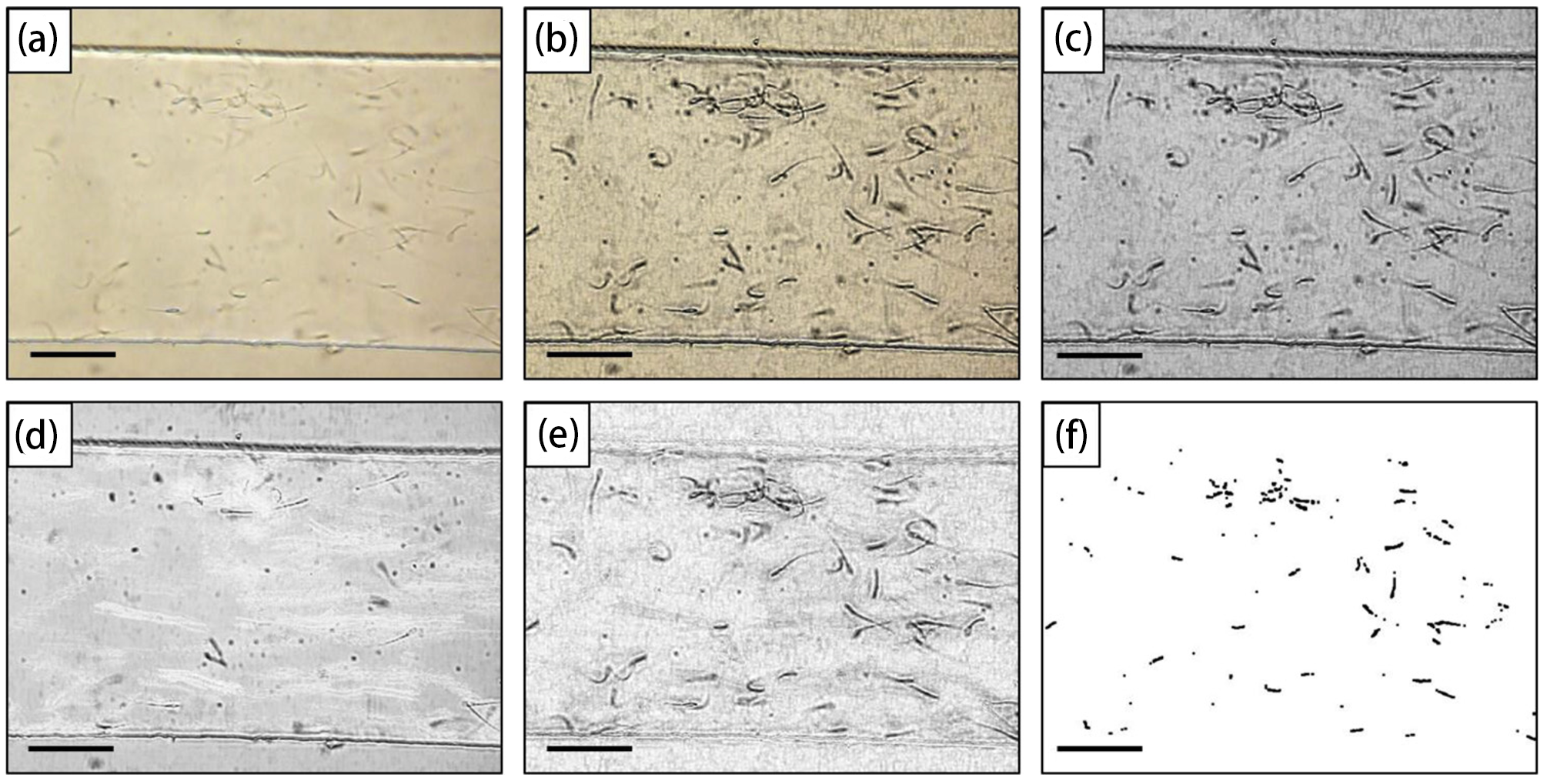}
\caption{ The processing result for each step of proposed method in \cite{elsayed-2015-DCSA}. The figure corresponds to Fig.2 in the original paper.}
\label{fig:CDSA}
\end{figure}

In~\cite{elsayed-2015-DCSA}, different algorithms are integrated for a sperm segmentation 
task, including grayscale conversion, background identification, background subtraction, 
binarization, and binary morphology approaches. First, an adaptive histogram equalization method is applied to enhance local contrast of input image. After that, for better processing, image enhanced is transformed into gray-scale image. Third, the background generated is respectively employed for all frames based on a subtraction operation to eliminate image noise. Maximum entropy algorithm achieves the best performance compared to other binarization methods in image binarization experiment. Therefore, maximum entropy algorithm is applied for image binarization at last. Fig.\ref{fig:CDSA} shows the processing result.

\par
\begin{figure}[htbp!]
\centering
\includegraphics[width=0.98\linewidth]{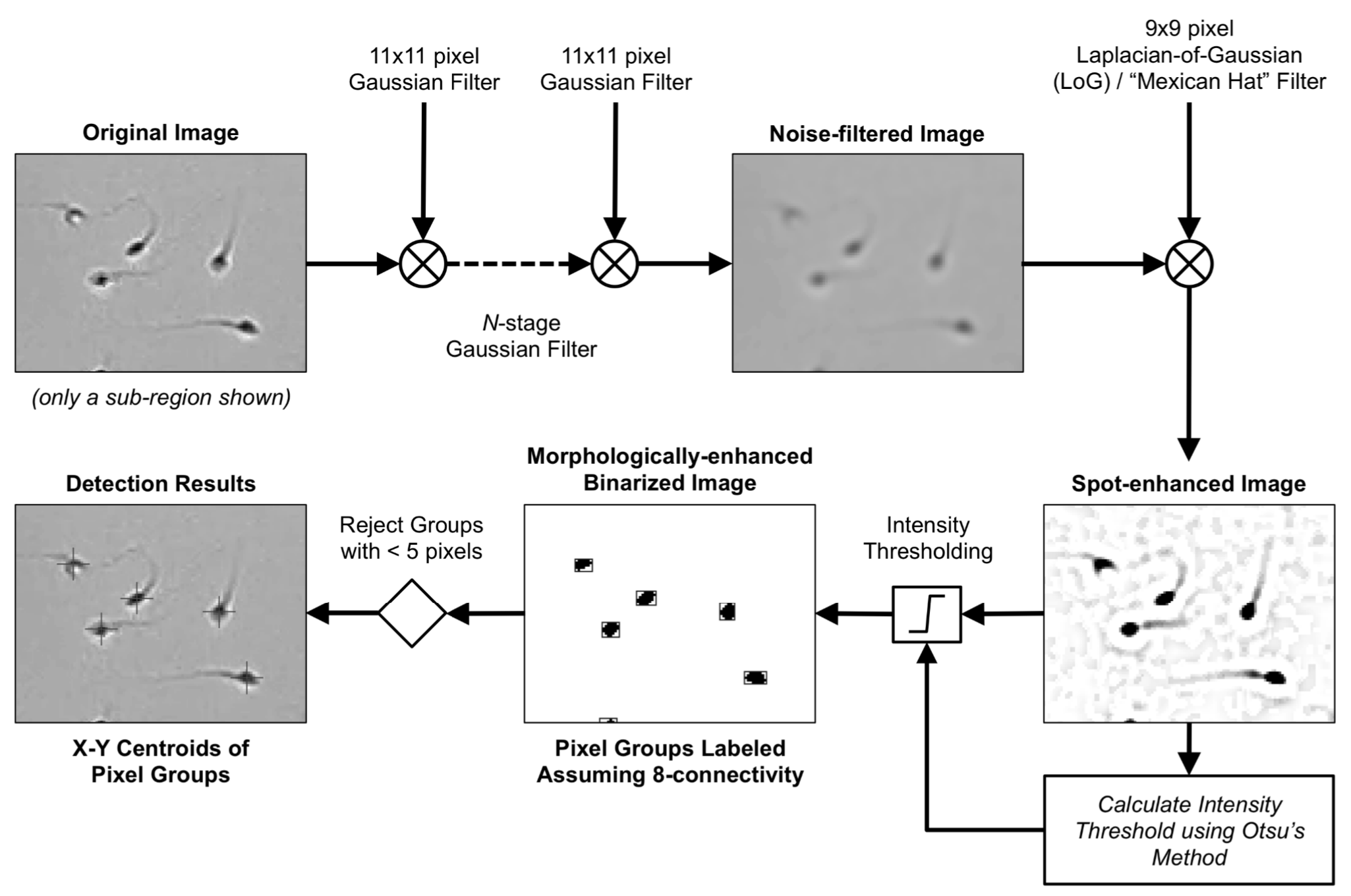}
\caption{ The overview of proposed method mentioned in \cite{urbano-2016-ATMA}. The figure corresponds to Fig.1(b) in the original paper.}
\label{fig:AT}
\end{figure}

The processing steps of proposed method in \cite{urbano-2016-ATMA} is illustrated in Fig.\ref{fig:AT}. First of all, a Gaussian filter is performed several times for eliminating noise and optimizing edge. Second, Laplacian-of-Gaussian (LoG) (``Mexican-hat'') filter is applied for enhancing the contrast between the object and its surrounding background. Based on this step, a spot-enhancement image is obtained. After that, the processed image is binarized by employing Otsu threshold. At last, a series of morphological operations are performed for binary image to improve detection accuracy. In the final processed image, only objects with less than 5 pixels are considered as sperm heads. Experiment results on prepared video frames suggest that the proposed method yields a high accuracy of $95\%$.

\par
\begin{figure}[htbp!]
\centering
\includegraphics[width=0.75\linewidth]{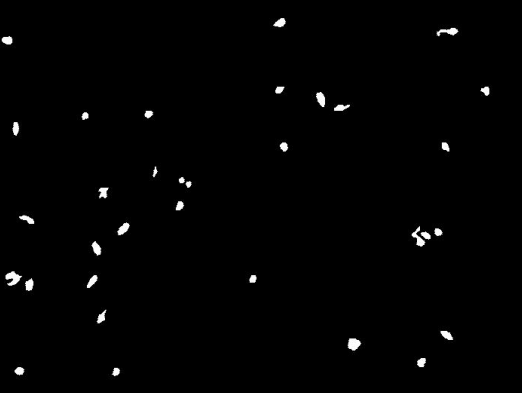}
\caption{ The segmentation result of proposed method in \cite{li-2020-FFFD}. The figure corresponds to Fig.4 in the original paper.}
\label{fig:FF}
\end{figure}

In \cite{li-2020-FFFD}, an automatic threshold segmentation method is proposed for sperm detection. The key idea of proposed method is to set a threshold named $T$ in advance. The input image is then binarized based on threshold $T$. Pixels with a pixel value above $T$ are setted to one. Other pixels are setted to zero. Fig.\ref{fig:FF} exhibits a result based on this method.

\subsection{Other Methods}
\label{ss:Detection:other}

\begin{figure}[htbp!]
\centering
\includegraphics[width=0.75\linewidth]{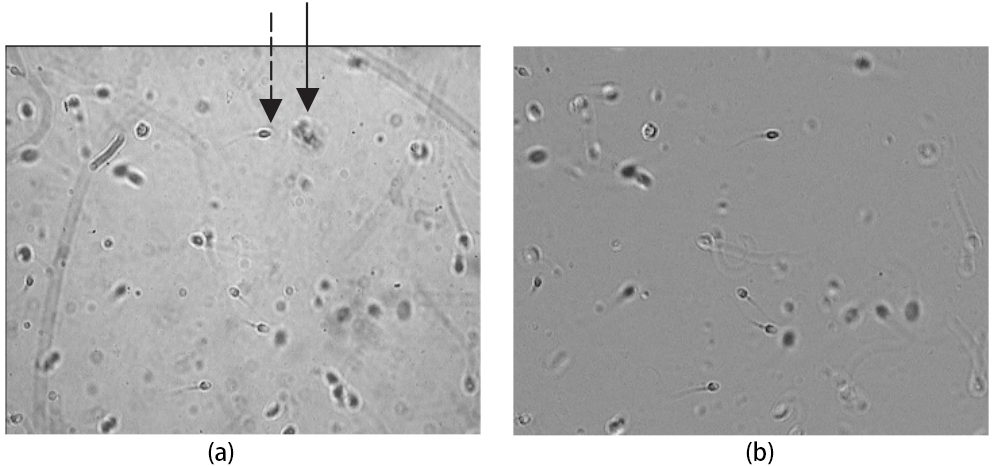}
\caption{ The enhancement result of proposed method in \cite{nafisi-2005-TMAS}. The figure corresponds to Fig.1 in the original paper.}
\label{fig:TM}
\end{figure}

\paragraph{\textbf{Semi-automatic Methods}}
In \cite{nafisi-2005-TMAS}, an image enhancement method with two steps is applied for sperm detection. First, the position invariant parts of the continuous frame are removed. Second, parts with fuzzy boundaries are eliminated.  After that, sperms in enhanced image are manually labelled by professional operator. All the labelled sperms is prepared for template matching method. The result after image enhancement is shown in Fig.~\ref{fig:TM}(b).

\par
In \cite{nosrati-2015-TSSS}, the professional software ImageJ \cite{ImageJ-2020-WOI} is employed for tracking sperm head from continuous frames of images. In addition, sperm position is determined by center of its head.

\begin{figure}[htbp!]
\centering
\includegraphics[width=0.75\linewidth]{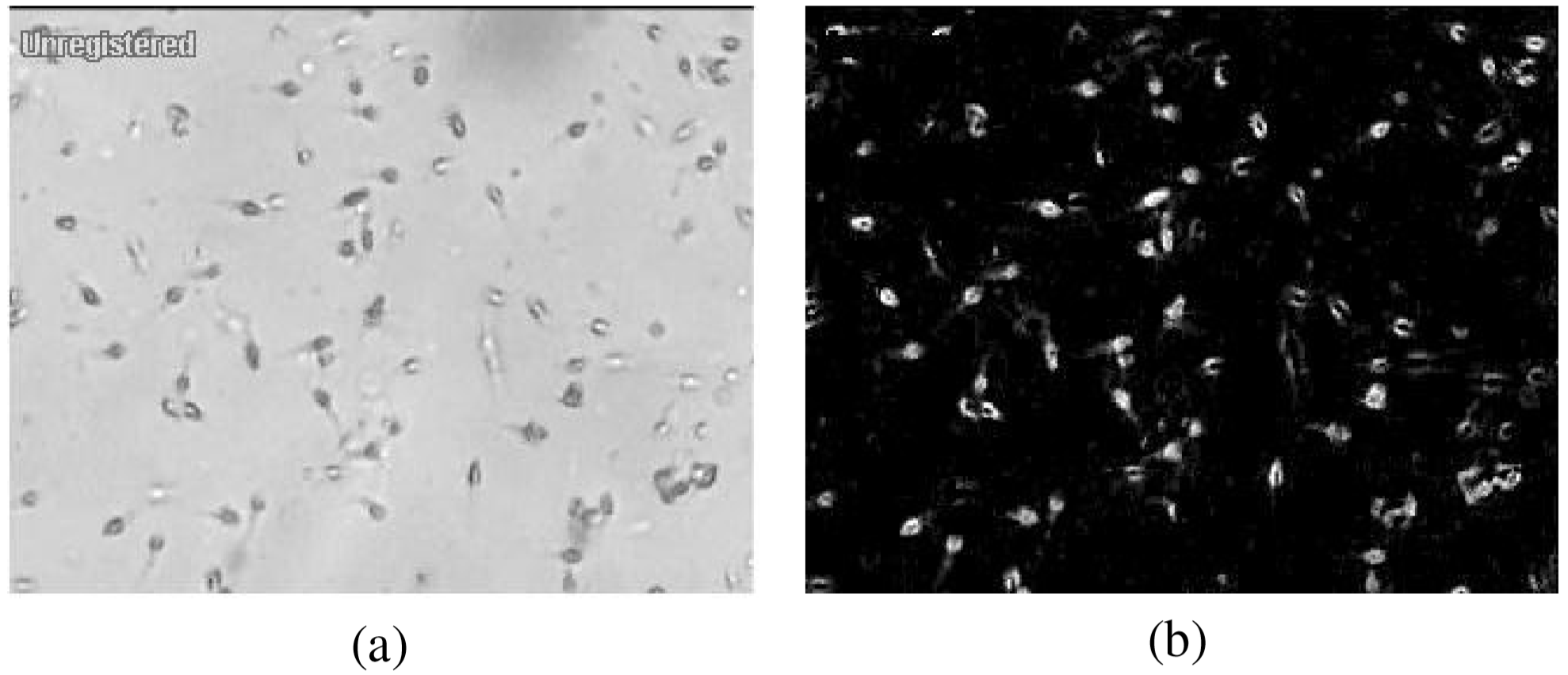}
\caption{The filtering result of proposed method in \cite{ravanfar-2011-LCSD}. The figure corresponds to Fig.1 in the original paper.}
\label{fig:LC}
\end{figure}

\paragraph{\textbf{Filtering Methods}}
In \cite{ravanfar-2011-LCSD}, sperm is distinguished from the background based on morphological filters. The key idea of this method is designing a structure element that is similar in size and shape to the sperm head to filter all images by employing Top-hat method. In addition, an open filter is applied for removing objects with areas smaller than the standard sperm head size. After that, background is removed by employing a special median filter. The detection results are shown in Fig.\ref{fig:LC}.
\par
In \cite{nurhadiyatna-2014-CIMD}, an enhancement Gaussian Mixture Model (GMM) called GMMHF is proposed for sperm detection. Compared with initial GMM, the proposed one is improved based on Hole Filling method. The key idea of GMMHF is grouping all pixels into foreground and background, and calculating the partition probability of each pixel. The 
result binaries of foreground and background are obtained by a large number of probability 
calculations. In this work, the researchers compare other methods with GMMHF. The experiment 
concluded that, although it is not the most accurate detection algorithm, it requires a 
relatively small amount of computation. 
\par

\begin{figure}[htbp!]
\centering
\includegraphics[width=0.75\linewidth]{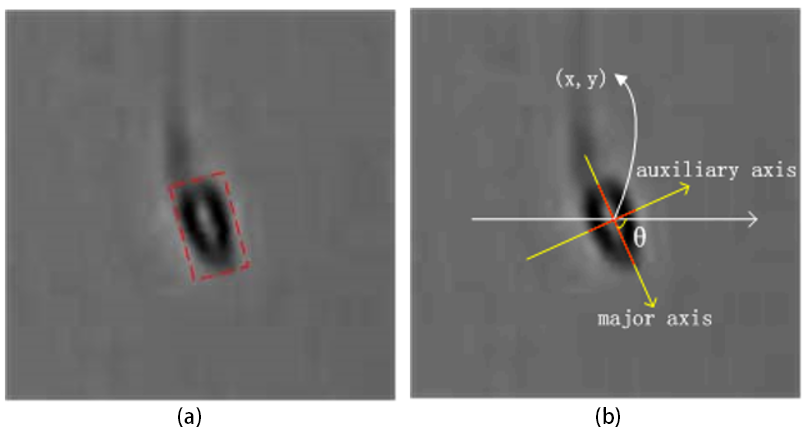}
\caption{The representation result of proposed method in \cite{zhou-2009-EMSP}. The figure corresponds to Fig.1 in the original paper.}
\label{fig:EM}
\end{figure}

\paragraph{\textbf{Shape Fitting Methods}}
The key idea of sperm detection in \cite{zhou-2009-EMSP} is creating a rectangle region to surround sperm. Parameters of the rectangle are correspondent to the current state of sperm. An example of rectangle region generated is shown in Fig.\ref{fig:EM}. Among this region, ($x,y$) means the center while $\theta$ means the orientation of the sperm in the image.

\par
\begin{figure}[htbp!]
\centering
\includegraphics[width=0.75\linewidth]{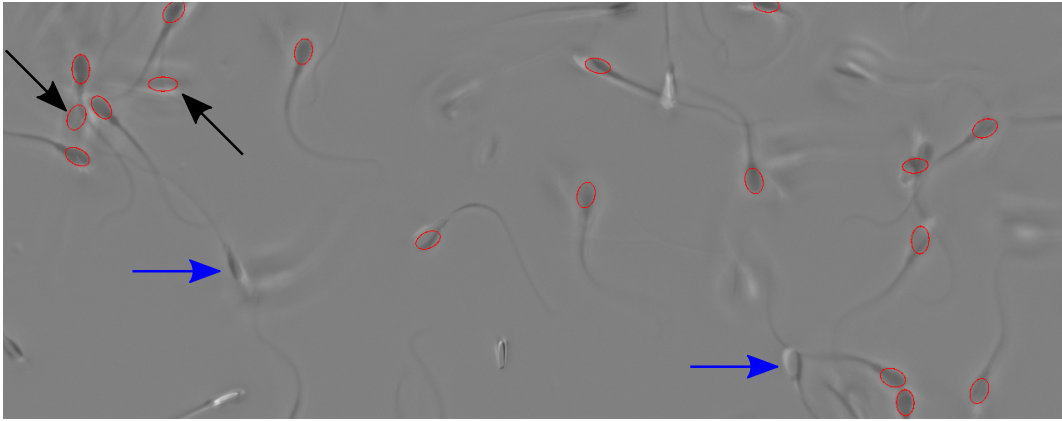}
\caption{ The detection result of proposed method in \cite{yang-2014-HTFT}. The figure corresponds to Fig.3 in the original paper.}
\label{fig:HT}
\end{figure}

In \cite{yang-2014-HTFT}, a model based method is proposed for detecting and localizing sperm head. Sperm head is replaced by an ellipse in this model, which contains five parameters  ($x, y, a, b, \phi$). ($x,y$)  represents the coordinate of the center point. $a$ and $b$ represents respectively the lengths of the long and short axes. $\phi$ is the angle between the major axis and the horizontal direction. Based on this model, the task of sperm head detection is converted into ellipse-shape objects detection. For ellipse-shape object detection, an improved multiple birth and cut (MBC) is employed. Fig.\ref{fig:HT} shows some detection results.

\paragraph{\textbf{Unsupervised Learning-based Methods}}
In~\cite{berezansky-2007-STHS}, it uses an approach called Spatio-Temporal Segmentation 
to detect sperm. Segmentation of video sequences by modeling the video data as 
multidimensional space-time data is applied in this work. It integrates $k$-means, GMM, 
mean shift and other segmentation detection and background separation algorithms.

In~\cite{z-2006-RATA}, the use of optical capture method for sperm detection is applied, 
where computer methods are partly used to assist manual work.

\subsection{Summary}
\label{ss:Detection:summary}
A summary of detection (localization) works reviewed is given in Tab.~\ref{tab:tabledetection}. 
We can see that many teams adopt self-developed detection algorithms to locate sperm and 
get good experimental results. Methods based on threshold are widely applied in sperm detection. Compared with other methods, methods based on threshold are simpler and easier to understand. In addition, methods based on threshold consist of many famous methods, such as Otsu threshold, local adaptive threshold. These alternatives 
give wider space for detecting and make threshold-based algorithm suitable to many different 
detection challenges that present on sperm microscopic video. 
\begin{table*}[htbp!]
\centering
\caption{A summary of sperm detection (localization) works in CASA field.}
\label{tab:tabledetection}
\renewcommand\arraystretch{1.5}
\setlength{\tabcolsep}{12pt}
\resizebox{11.6cm}{!}{
\begin{tabular}{@{}ccccc@{}}
\toprule
Methods                      & Technical Details                             & Time & Research Groups                  & References \\ \midrule
\multirow{15.5}{*}{Thresholding Based} & Thresholding detection     & 1988 & Serdia O Mack          & ~\cite{Mack-1988-QSPM}  \\ \cmidrule(l){2-5}
                            & Thresholding detection              & 1998 & Michel Kraemer         & ~\cite{Kraemer-1998-FIHS}  \\ \cmidrule(l){2-5}                             
                            & Otsu                                & 2005 & Vahid Reza Nafisi      & ~\cite{nafisi-2005-SIUE}  \\ \cmidrule(l){2-5} 
                            & Thresholding detection              & 2010 &
VS Abbiramy          & ~\cite{Abbiramy-2010-SDCT}  \\ \cmidrule(l){2-5}  
                            & Otsu adaptive thresholding          & 2010 & Clement Leung          & ~\cite{leung-2010-DTLC}  \\ \cmidrule(l){2-5} 
                            & Thresholding detection              & 2010 & Mathew James Tomlinson & ~\cite{tomlinson-2010-VNCS}  \\ \cmidrule(l){2-5} 
                            & Otsu adaptive thresholding          & 2011 & Zhe Lu                 & ~\cite{lu-2011-robotic}  \\ \cmidrule(l){2-5} 
                            & Thresholding detection              & 2012 & Jun Liu                & ~\cite{liu-2012-QALB}  \\ \cmidrule(l){2-5} 
                            & Otsu criterion                      & 2014 & Giuseppe Di Caprio     & ~\cite{di-2014-4dTC}  \\ \cmidrule(l){2-5} 
                            & Optimum binarization                & 2015 & Mohamed Elsayed        & ~\cite{elsayed-2015-DCSA}  \\ \cmidrule(l){2-5} 
                            & Otsu                                & 2016 & Leonardo F Urbano      & ~\cite{urbano-2016-ATMA}  \\ \cmidrule(l){2-5}
                            & Thresholding detection              & 2020 & Xialin Li         & ~\cite{li-2020-FFFD}  \\ \midrule 
\multirow{9.5}{*}{Others}     & Semi-automatic Methods              & 2005 & Vahid Reza Nafisi      & ~\cite{nafisi-2005-TMAS}  \\ \cmidrule(l){2-5}
                            & Unsupervised Learning based Methods & 2007 & MichaelBerezansky      & ~\cite{berezansky-2007-STHS}  \\ \cmidrule(l){2-5} 
                            & Shape Fitting Methods               & 2009 & Xiuzhuang Zhou         & ~\cite{zhou-2009-EMSP}        \\ \cmidrule(l){2-5} 
                            & Filtering Methods                   & 2011 & Mohammad R Ravanfar    & ~\cite{ravanfar-2011-LCSD}  \\ \cmidrule(l){2-5} 
                            & Filtering Methods                   & 2014 & Adi Nurhadiyatna       & ~\cite{nurhadiyatna-2014-CIMD}  \\ \cmidrule(l){2-5} 
                            & Shape Fitting Methods               & 2014 & H-F Yang               & ~\cite{yang-2014-HTFT}  \\ \cmidrule(l){2-5} 
                            & Semi-automatic Methods              & 2015 & Reza Nosrati           & ~\cite{nosrati-2015-TSSS} \\ \bottomrule
\end{tabular}}
\end{table*}
\section{Sperm Tracking}
\label{s:Tracking}

After enhancing image quality and sperm detection, as the second step of CASA technology, sperm tracking algorithm plays a decisive role in the quality of the analysis results. The sperm has a lot of different movement patterns in video, including fast and slow linear motion, fast and slow bending motion. This means the position of the same sperm may change differently in different frames of video. Therefore, sperm tracking is a difficult task in high-dimensional state space \cite{zhou-2009-EMSP}.
\par
From the related works we review, it can be seen that different research teams have different 
research interests in sperm tracking, but from an application view, there are mainly two 
types of tracking: The sperm head region of interest (SHROI) and the sperm tail region of 
interest (STROI). Visual tracking, as a hot research area with great potential, is widely studied in recent years. There are several common tracking algorithms in this field, 
such as Particle filter (PF) algorithms, the Maximum Intensity Region (MIR) algorithms, 
Mean Shift (MF) algorithms, the Optical Flow (OF) algorithms. There are also self-developed 
tracking algorithms by some research teams. Next, we review the applications of tracking 
technology in the field of sperm microscopic video analysis.

\subsection{Head Tracking}
\label{ss:Tracking:head}
In \cite{Abbiramy-2010-SDCT}, first frame of prepared video is selected for setting a baseline. In addition, based on sperm detection processing, numbers of sperms in all frames are respectively calculated and stored. Considering that sperm has a comparative low speed, regions surrounding sperm are focused on. Therefore, it is not necessary to spend a lot of time and effort to analyse the entire image.At the beginning of sperm tracking, frames are divided into many regular regions based on centroid of objects. When objects detected have similar shape and size with reference object in first frame, It are regarded as sperm. Then, taking the centroid position of the first frame as the starting point, the distance of centroid movement is calculated. 
\par
\begin{figure}[htbp!]
\centering
\includegraphics[width=0.98\linewidth]{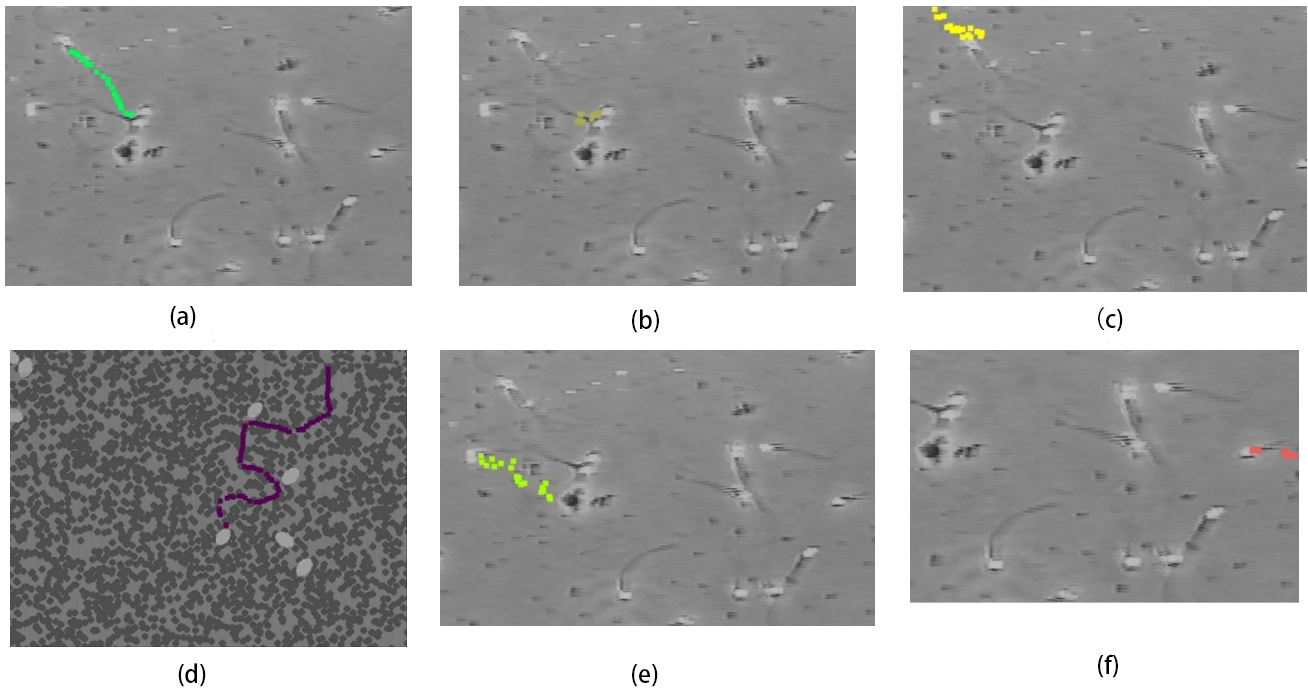}
\caption{The tracking result of proposed method in \cite{berezansky-2007-STHS}. The figure corresponds to Fig.5 in the original paper. }
\label{fig:ST}
\end{figure}

In \cite{berezansky-2007-STHS}, a 4-dimensional model consisting of two spatial coordinates, one time coordinate and one optical flow vectors direction is proposed. Based on proposed model, the problem of incorrect path reconstruction caused by collision is potential to be solved. First, an filtering is applied for generating pixel set of detected sperm. Then, pixels detected are represented by optical flow vectors. After processing all frames based on above method, a mean shift clustering method is employed for clustering all features obtained . At last, the correspondence between above clusters and initial path is analysed. 
The experiment results are shown in Fig.\ref{fig:ST}, and this reference calculates the moving speed of the sperm.

\par
\begin{figure}[htbp!]
\centering
\includegraphics[width=0.98\linewidth]{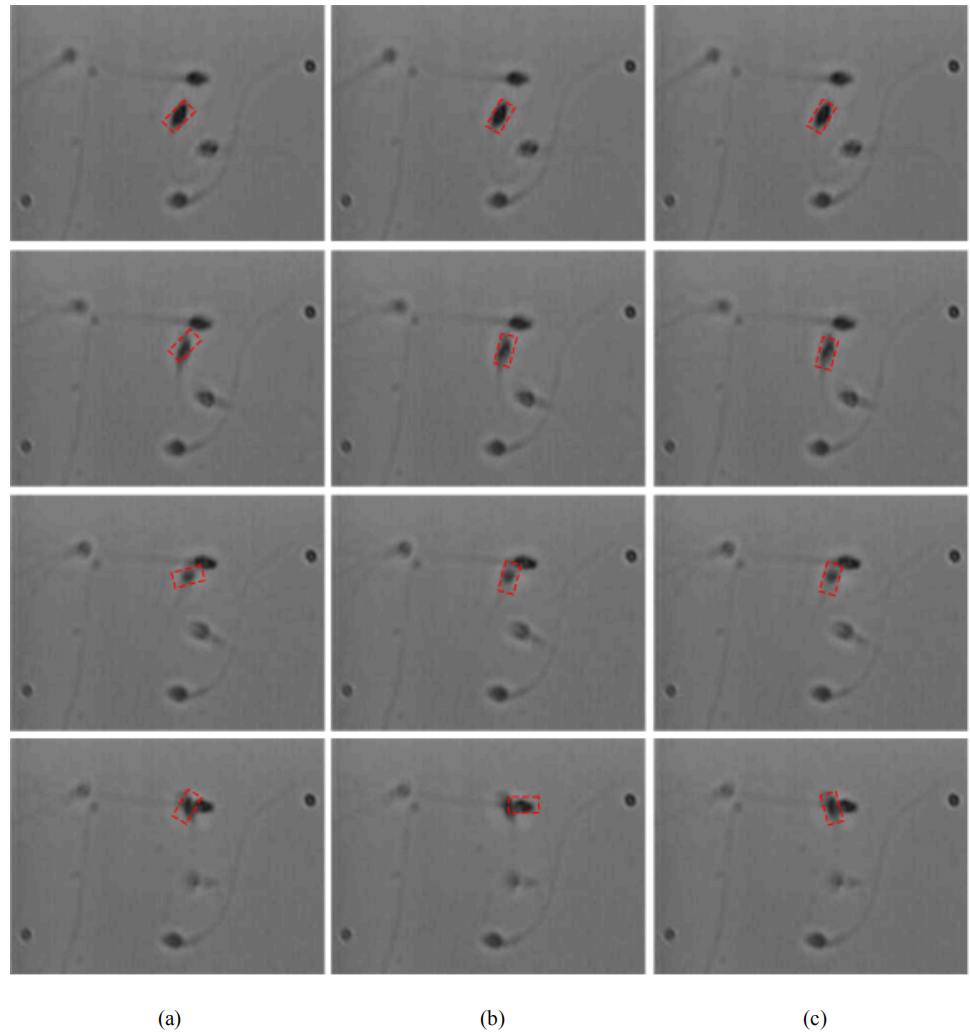}
\caption{The tracking results of proposed method in \cite{zhou-2009-EMSP}. The figure corresponds to Fig.3 in the original paper.}
\label{fig:EM1}
\end{figure}

In \cite{zhou-2009-EMSP}, an automatic method combining mean shift and a particle filter is proposed for tracking sperm. Sperm tracking using particle filters alone has the disadvantage of being too inefficient. Sperm tracking using particle filters alone has the disadvantage of falling into local optimal mode easily. In addition, both two methods fail to achieve efficiently tracking for sperms with random movement. Therefore, by combining this two methods, the proposed method can achieve rapid and accurate sperm tracking. The results of this algorithm are compared with  results of particle filter and mean shift respectively, seen in Fig.\ref{fig:EM1}.

\par
In \cite{tomlinson-2010-VNCS}, a multi-target tracker by employing tracking method mentioned in \cite{khan-2004-an} is proposed. Experiment shows that the proposed method is able to track all sperms appeared in a fixed region. In addition, with the assistance of operator, the tracking operation can be continued for the other one after processing a sperm video.
\par
\begin{figure}[htbp!]
\centering
\includegraphics[width=0.6\linewidth]{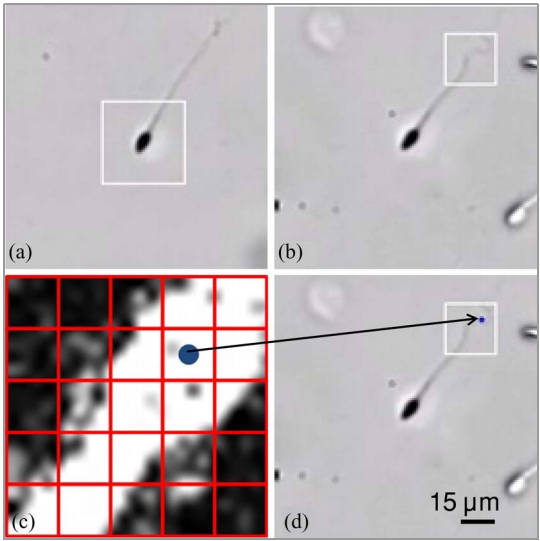}
\caption{ The tracking result of proposed method in \cite{lu-2011-robotic}. The figure corresponds to Fig.5 in the original paper.}
\label{fig:RI1}
\end{figure}

In \cite{leung-2010-DTLC,lu-2011-robotic}, an automatic sperm head tracking method is proposed. First step is locating sperm head based on analysing image contour. A $40 \times 40$ pixels region is then creating to surround sperm head, center  of which is at the same position with center of sperm head. At last, all frames in prepared video are performed by the same operations to achieve sperm tracking. In addition, to make the object always in the center of the monitor area, the distance between center of the monitor area and sperm head is calculated. The distance data obtained is then employed by Proportion Integral Differential (PID) controller, which is able to make sperm in the center by moving view field. Fig.\ref{fig:RI1} illustrates the result of sperm tracking based on proposed method. 

\par
\begin{figure}[htbp!]
\centering
\includegraphics[width=0.75\linewidth]{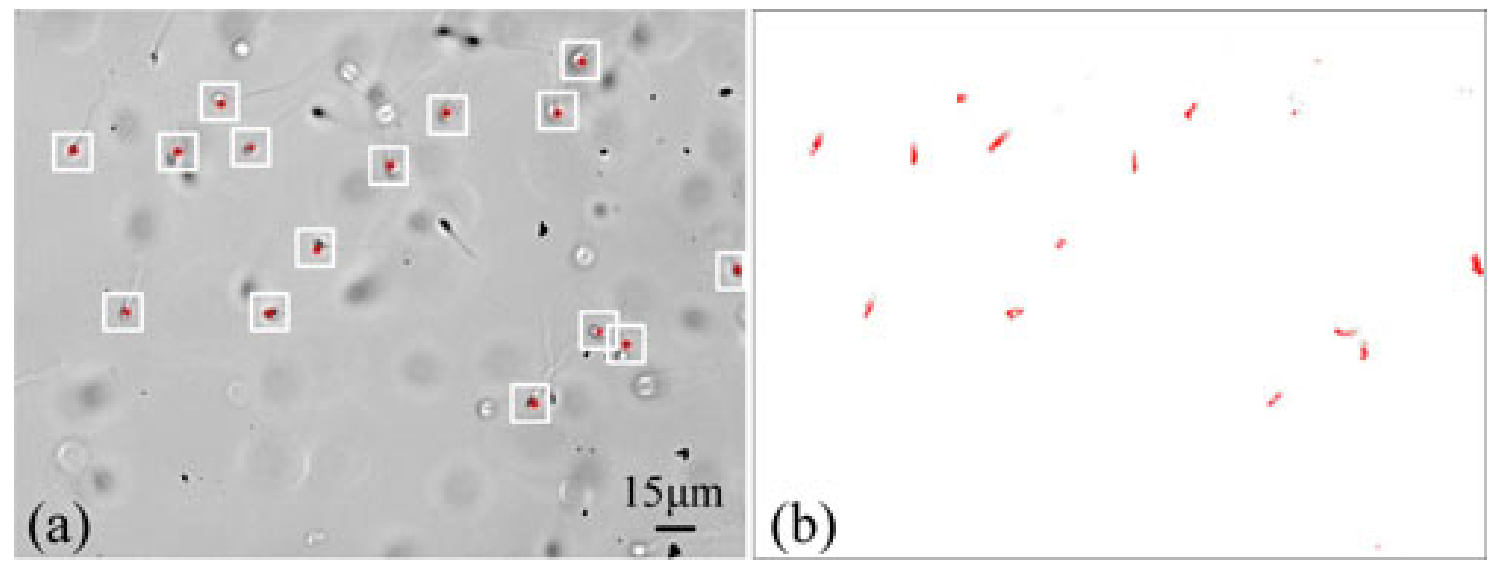}
\caption{ The tracking results of proposed method in \cite{liu-2012-QALB}. The figure corresponds to Fig.2 in the original paper.}
\label{fig:QA}
\end{figure}

In \cite{liu-2012-QALB}, a novel tracking methods is proposed for automatically tracking multiple moving sperm heads in the field of view at the same time. First, a subtraction operation is performed between two consecutive frames. Based on this step, a rough contour of sperm is obtained, which is then optimized by employing a fixed threshold value of 60.  As the sperm video is processed frame by frame, the new captured contour of moved sperm replaces the previously captured contour in each frame of the video while the old contour is disappeared over time. At last, with the corresponding parameter settings, each contour exists for 0.5 seconds, and the color of the contour gradually lightens over time. Therefore, according to the sperm head contour traces of different shades of color, the movement route of sperm is obtained. In addition, to reduce the influence of noise and Brownian motions, several morphological operations are applied. The sperm tracking result is exhibited in Fig.\ref{fig:QA}.

\par
In \cite{nafisi-2005-TMAS}, specific sperms are first chosen by professional operator. Center of selected sperm's head is labelled manually. After that, method mentioned in \cite{teifoory-2002-NMSS} is employed for finding out the region that surrounds sperm head. Based on this method, edge of sperm head can be efficiently detected and the minimum box that surrounds sperm head is generated.
\par
\begin{figure}[htbp!]
\centering
\includegraphics[width=0.75\linewidth]{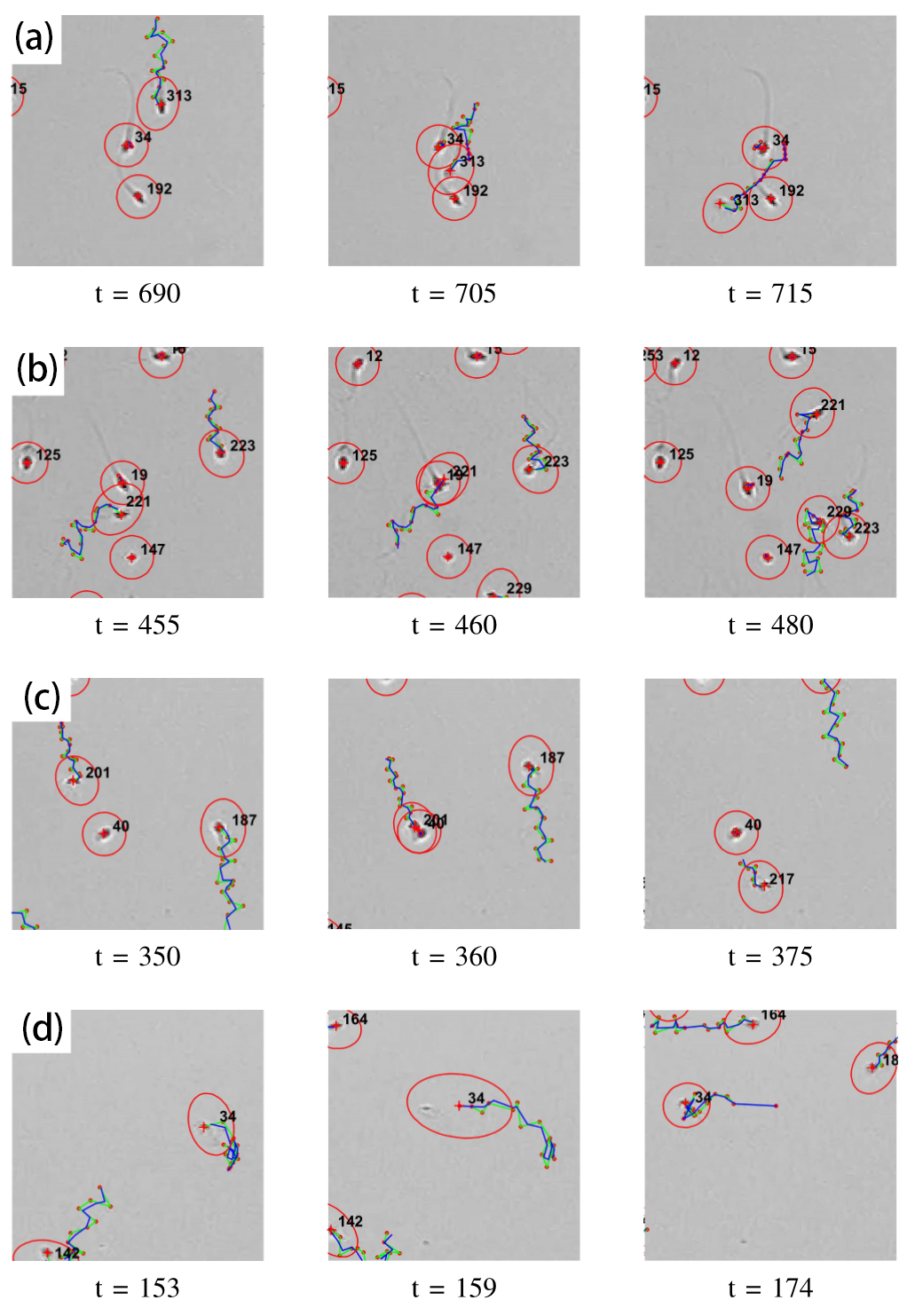}
\caption{The tracking results of proposed method in \cite{urbano-2016-ATMA}. The figure corresponds to Fig.4 in the original paper.}
\label{fig:AT1}
\end{figure}

In \cite{urbano-2016-ATMA}, a joint probabilistic data association filter (JPDAF) method is proposed based on Matlab platform. Compared to original PDAF, the novel one comprehensively considers all the event probabilities associated with sperm, which is NP-hard \cite{crouse-2010-TGIP,romeo-2012-AFCA}. A probabilityweighted sum is applied by JPDAF to update sperm path. In addition, Kusiak’s method \cite{kusiak-1987-AECI} is employed in processing paths and measurements obtained to achieve faster processing. Moreover, Cox’s adaptation method \cite{blackman-1999-DAAO} of Kusiak’s is employed to reduce the amount of computation and computation time by selecting most highly probable joint association events. Since sperm cells can exhibit abrupt maneuvers, each independent Kalman filter uses a continuous white noise acceleration (CWNA) target motion model with an adaptive process noise covariance. Fig.\ref{fig:AT1} exhibits tracking results based on proposed method.
\par

\begin{figure}[htbp!]
\centering
\includegraphics[width=0.7\linewidth]{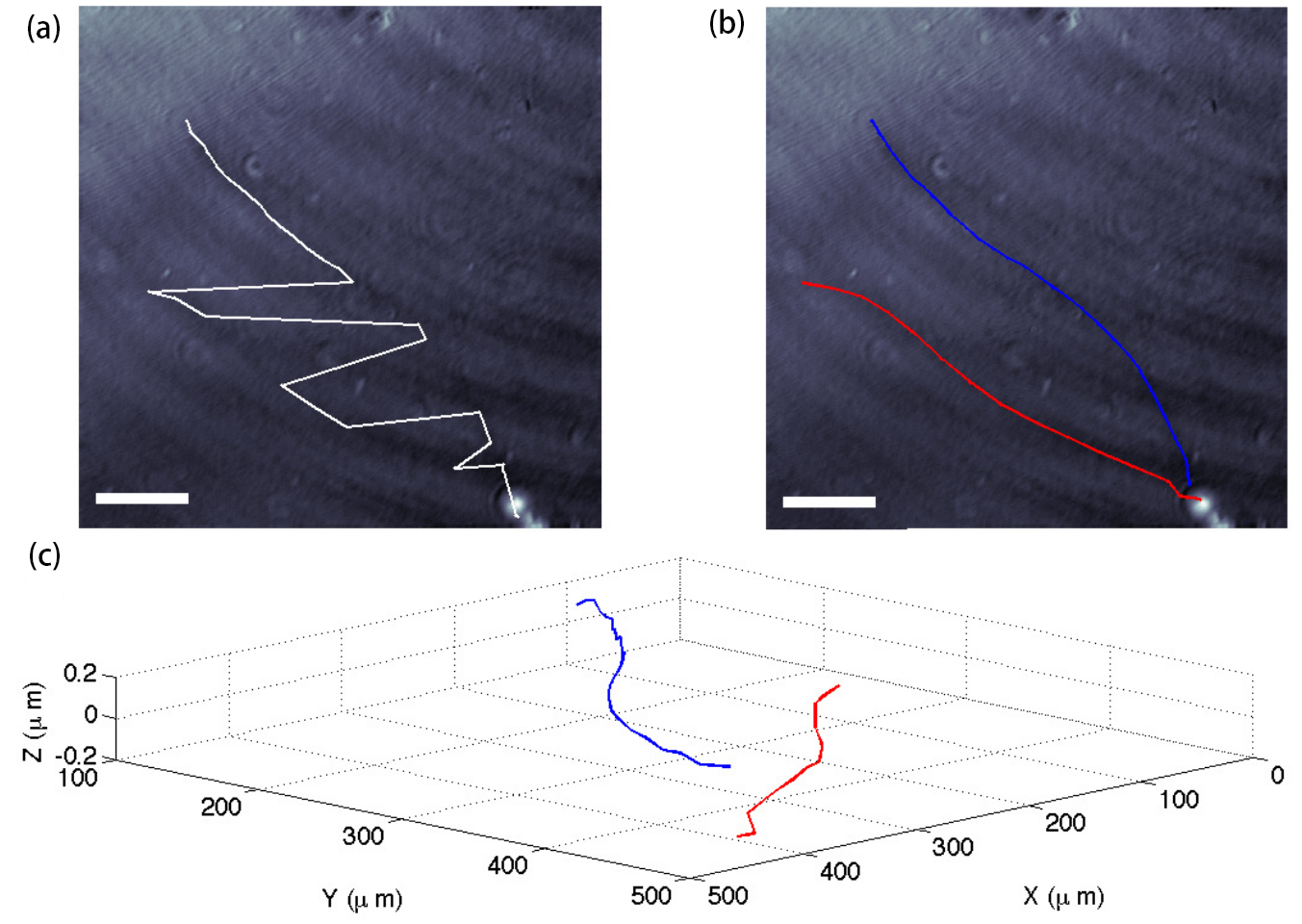}
\caption{ The tracking results of proposed method in \cite{di-2014-4dTC}. The figure corresponds to Fig.5 in the original paper.}
\label{fig:4D}
\end{figure}

In \cite{di-2014-4dTC}, a lot of computation time and effort is saved by employing proximity criterion. Specifically, according to the position of sperm in the $n$-th frame, unreasonable regions in the $(n+1)$-th frame are not considered. The tracking result with the assistance of proximity criterion is shown in Fig.\ref{fig:4D}. Although it appears from Fig.\ref{fig:4D}(b) that the paths of the two sperm seem to cross. In fact, they are on different planes, seen in Fig.\ref{fig:4D}(c).In addition, proximity criterion is employed in $Z$ plane to achieve a tracking result with higher reliability. Based on this idea, objects with low contrast of phase map can be tracked.
\par
\begin{figure}[htbp!]
\centering
\includegraphics[width=0.78\linewidth]{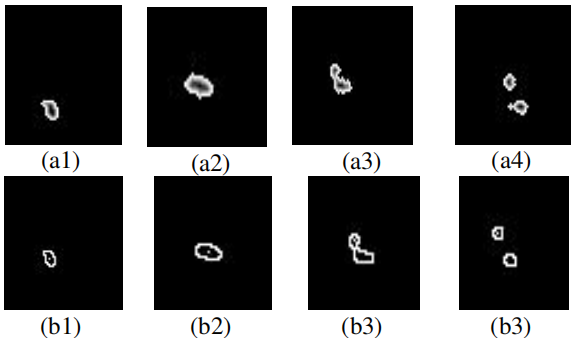}
\caption{ The tracking results of proposed method in \cite{ravanfar-2011-LCSD}. The figure corresponds to Fig.2 in the original paper.}
\label{fig:LC1}
\end{figure}

In \cite{ravanfar-2011-LCSD}, a particle filter algorithm is applied to track the sperm. 300 particles is prepared for tracking one sperm. In addition, special sperm is hard to separate with others only based on velocity data when collision happens. Therefore, watershed method is employed for separating the special one with other sperms when collision happens. The Fig.\ref{fig:LC1} shows the results of the treatment of sperm collision.
\par

\begin{figure}[htbp!]
\centering
\includegraphics[width=0.9\linewidth]{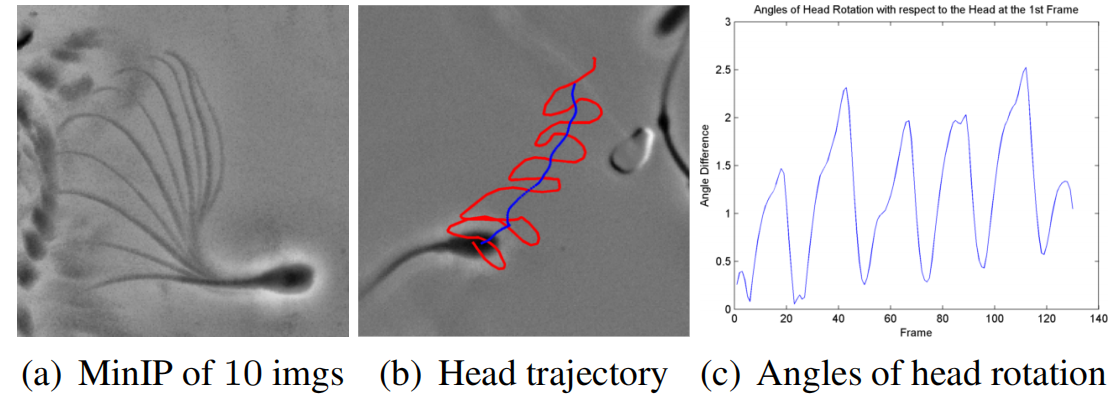}
\caption{The tracking results of proposed method in \cite{yang-2014-HTFT}. The figure corresponds to Fig.4 in the original paper.}
\label{fig:HT2}
\end{figure}

In \cite{yang-2014-HTFT}, a reliable method based on two frames in a row is proposed for tracking sperm in video. The two frames in a row is regarded as a set of sample. For each sample, a sub-frame surrounding sperm head is generated in the previous frame, which is then matched to the next frame. The movement of sperm head is obtained based on the matching step, which is employed to generate a sub-frame for the next frame. The whole processing steps consisting of two stages is based on block matching method. The first stage is finding out the most suitable region by applying a similarity criterion. After that, the second stage is selecting most suitable global rigid transformation for all correspondences gained above. Fig.\ref{fig:HT2} shows some head tracking results.

\par
The task of sperm tracking presents two major difficulties. For one thing, sperm movement is relatively rapid and random. Most sperm, on the other hand, are similar in size and shape. In \cite{jati-2015-HSTU}, a sperm tracking method combining particle swarm optimization and smoothing stochastic approximate monte carlo is proposed. In addition, to save the effort of finding areas that contain sperm, a windowing method is employed.
\par
\begin{figure}[htbp!]
\centering
\includegraphics[width=0.75\linewidth]{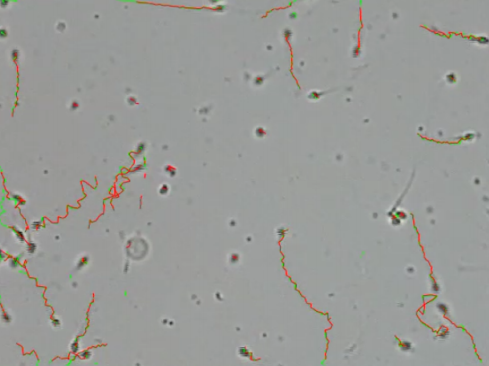}
\caption{The classification results of proposed method in \cite{li-2020-FFFD}. The figure corresponds to Fig.5 in the original paper.}
\label{fig:FF1}
\end{figure}

In \cite{li-2020-FFFD}, an automatic method combining computer technology and image processing technology is designed for sperm tracking. One challenge faced by the sperm tracking task is to determine the position of the same object in multiple consecutive frames of images. A $k$-nearest neighbor ($k$-NN) \cite{goldstein-1972-KNNC} is finally selected to deal this challenge. An tracking result based on $k$-NN is shown in Fig.\ref{fig:FF1}.

\subsection{Tail Tracking}
\label{ss:Tracking:tail}
\begin{figure}[htbp!]
\centering
\includegraphics[width=0.75\linewidth]{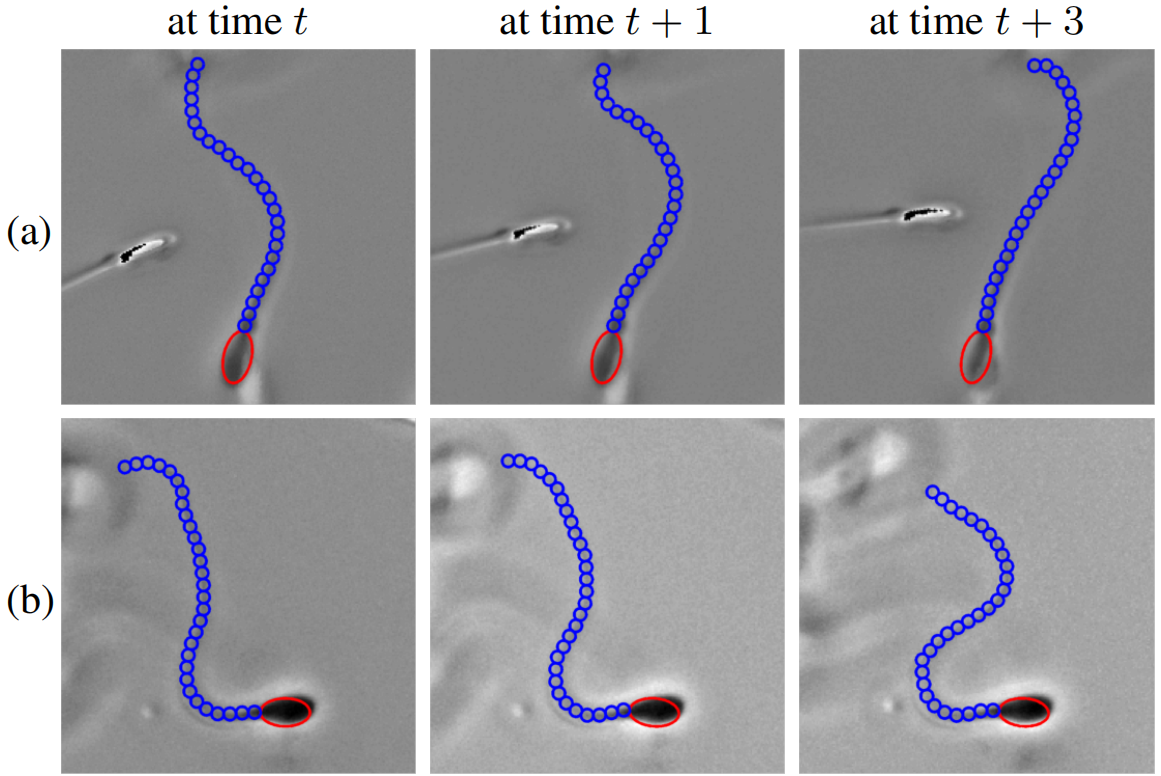}
\caption{The tracking results of proposed method in \cite{yang-2014-HTFT}. The figure corresponds to Fig.5 in the original paper.}
\label{fig:HT3}
\end{figure}

In \cite{yang-2014-HTFT}, a novel algorithm for tail tracking is proposed. Based on the proposed method, shapes of flagellar beat are able to be obtained. The key idea of proposed method is replacing flagellum with several circles. Fig.\ref{fig:HT3} shows some flagellum tracking results.

\subsection{Summary}
\label{ss:Tracking:summary}
From this chapter, we can see that even though the processing of microscopic video images 
is generally the same, there are various views and applications of the tracking algorithm 
for sperm movement trajectory in different research teams. A summary of tracking works 
reviewed is shown in Tab.~\ref{tab:tabletracking}. Among them are the distance formula 
algorithm based on mathematics, and the conventional optical flow algorithm, the classical 
particle filter algorithm. There is also a tracking algorithm developed by the research team. 
However, due to the different data adopted by each algorithm, there are differences in 
evaluation standards, and there is no unified evaluation standard for CASA problem in the 
world. Therefore, the universality of these algorithms also remains to be investigated. 
Even so, from the experimental results, each algorithm shows a good tracking effect. 
\begin{table*}[htbp!]
\centering
\caption{ A summary of sperm tracking works in CASA field.}
\label{tab:tabletracking}
\renewcommand\arraystretch{1.5}
\setlength{\tabcolsep}{12pt}
\resizebox{11.5cm}{!}{
\begin{tabular}{@{}ccccc@{}}
\toprule
Methods                      & Technical Details                             & Time & Research Groups                  & References \\ \midrule
\multirow{20.5}{*}{Head Tracking} & Template Matching Algorithm     & 2005 & Vahid Reza Nafifisi          & ~\cite{nafisi-2005-TMAS}  \\ \cmidrule(l){2-5}
                            & 4-dimensional model              & 2007 & Michael Berezansky         & ~\cite{berezansky-2007-STHS}  \\ \cmidrule(l){2-5}                             
                            & \begin{tabular}[c]{@{}c@{}}The Combination of Mean Shift \\ and Particle Filter\end{tabular}                                & 2009 & Xiuzhuang Zhou      & ~\cite{zhou-2009-EMSP}  \\ \cmidrule(l){2-5} 
                            & Distance methods                                                                                                                     & 2010 &
VS Abbiramy          & ~\cite{Abbiramy-2010-SDCT}  \\ \cmidrule(l){2-5}  
                            & Distance methods          & 2010 & Clement Leung          & ~\cite{leung-2010-DTLC}  \\ \cmidrule(l){2-5} 
                            & Khan's Markov chain Monte Carlo              & 2010 & Mathew James Tomlinson & ~\cite{tomlinson-2010-VNCS}  \\ \cmidrule(l){2-5} 
                            & Distance methods                                                                                                                 & 2011 & Zhe Lu                 & ~\cite{lu-2011-robotic}  \\ \cmidrule(l){2-5} 
                            & Particle Filter                                                                                                                      & 2011 & Mohammad R Ravanfar                & ~\cite{ravanfar-2011-LCSD}  \\ \cmidrule(l){2-5} 
                            & Motion History Image                                                                                                                         & 2012 & Jun Liu     & ~\cite{liu-2012-QALB}  \\ \cmidrule(l){2-5} 
                            & Proximity Criterion                                                                                                                    & 2014 & Giuseppe Di Caprio        & ~\cite{di-2014-4dTC}  \\ \cmidrule(l){2-5} 
                            & Co-register Successive Images                                                                                                                          & 2014 & H-F Yang      & ~\cite{yang-2014-HTFT}  \\ \cmidrule(l){2-5}
                            & \begin{tabular}[c]{@{}c@{}}Particle Swarm Optimization and Smoothing \\ Stochastic Approximate Monte Carlo\end{tabular}             & 2015 & Grafifika Jati & ~\cite{jati-2015-HSTU}  \\ \cmidrule(l){2-5}                             
                            & \begin{tabular}[c]{@{}c@{}}Joint Probabilistic Data \\ Association Fifilter\end{tabular}                                & 2016 & Leonardo F Urbano      & ~\cite{urbano-2016-ATMA}  \\ \cmidrule(l){2-5}                                                         
                            & \textit{k}-NN                                                                                         & 2020 & Xialin Li         & ~\cite{li-2020-FFFD}  \\ \midrule 
\multirow{1}{*}{Tail Tracking}     & Shape Extraction              & 2014 & H-F Yang      & ~\cite{yang-2014-HTFT}  \\  
\bottomrule
\end{tabular}}
\end{table*}
\section{The Future of Computer-aided Sperm Analysis}
\label{s:future}

\subsection{Methods Analysis}
\label{ss:future:Analysis}

Through the previous summary, we can find that in the target detection stage, most 
teams adopt more traditional image segmentation methods: threshold detection, Otsu, 
and Otsu adaptive threshold~\cite{Dhal-2019-ASO,Dhal-2020-NOA}. Some research teams also choose 
graphic fitting and semi-automatic detection. However, in terms of the detection effect, 
no matter which detection method can obtain more accurate detection of sperm cells, 
each method has achieved satisfying experimental results. From this view, CASA technology 
has no great difficulty in sperm cell detection.

When it comes to tracking sperm motility, each team takes a different approach. There 
are clear and simple distance calculation methods, probabilistic tracking algorithms, 
and more traditional particle filtering algorithms. The relatively new shape extraction 
method has also achieved acceptable tracking effect in~\cite{yang-2014-HTFT}, but its 
effect depends partly on the image quality of the microscopic video. As a classical 
tracking technique, particle filter algorithm is also applied in the CASA system, but 
its tracking accuracy is also closely related to the number of particles. This leads 
to a drawback of particle filter algorithm, when the number of particles is enough, 
the calculation speed of the algorithm will be substantially reduced, at the same time, 
at the cost of a high requirement on the storage space of the system. This is an unavoidable 
problem for clinical devices. In~\cite{li-2020-FFFD}, $k$-NN is combined with deep learning. The convolution method is employed for features extraction from microscopic images, which greatly improves the accuracy of $k$-NN classification. After that, the researcher 
team uses distance calculations to track the movement of the same sperm. Feature 
extraction and target classification based on deep learning can identify and calculate 
the same sperm target more accurately for the tracking algorithm of distance operation. 
From Fig.~\ref{fig:FF1}, we can also see that the tracking effect of this algorithm 
is more similar to the real sperm movement track.

\subsection{Technical Limitations}
\label{ss:future:limitations}
With the development of computer related technology and the improvement of related tracking algorithm, CASA system will provide clearer images for sperm as well as make the sperm movement trajectory depicted by the CASA system closer and closer to the real sperm movement trajectory. There are many factors 
influencing the accuracy of CASA results. On a technical level, the main factors generally 
include: the optical system for capturing video, frame rate, algorithms (detection and 
tracking algorithms).

\begin{figure*}[htbp!]
\centering
\includegraphics[width=0.98\linewidth]{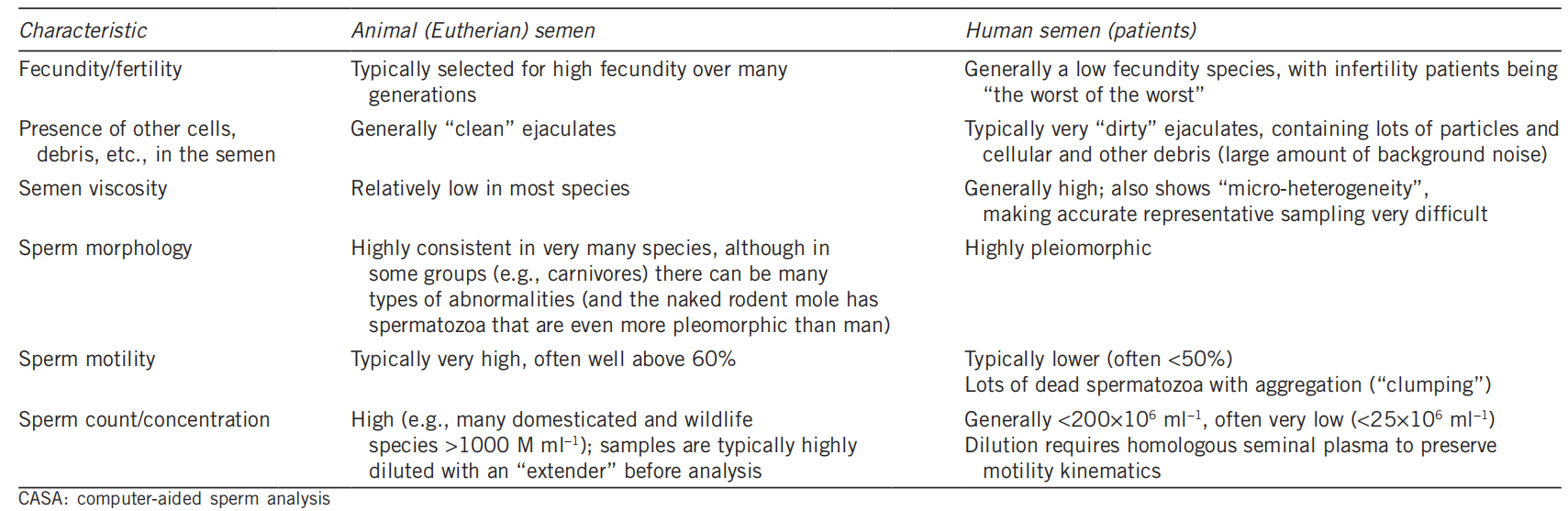}
\caption{ Biological factors that influence CASA analysis. This figure corresponds to Tab.~1 in the original paper~\cite{st-2015-TFCS}.}
\label{fig:TF}
\end{figure*}

For the problem of sperm trajectory tracking, the methods adopted in the early stage of image 
processing are roughly the same, because most of them adopt the means of computer graphics 
to detect and locate sperm. However, there are many kinds of tracking algorithms for target 
tracking. This leads to technical limitations for CASA, which focus on issues related to 
target tracking and fundamentally different ways of defining motion. Because of this, the CASA 
system has not been widely used in human medical clinical diagnosis. Relative to the original 
manual detection, CASA has some uncertainties. These factors may cause different systems to 
have different evaluation results for the same sperm sample. Nevertheless, the CASA system has 
been widely used in animal reproduction experiments and has achieved success. 
In~\cite{st-2015-TFCS}, it is attributed to the ``biological'' reason, and the main influencing 
factors are compared in the article, see in Fig.~\ref{fig:TF}. 

\begin{figure}[htbp!]
\centering
\includegraphics[width=0.75\linewidth]{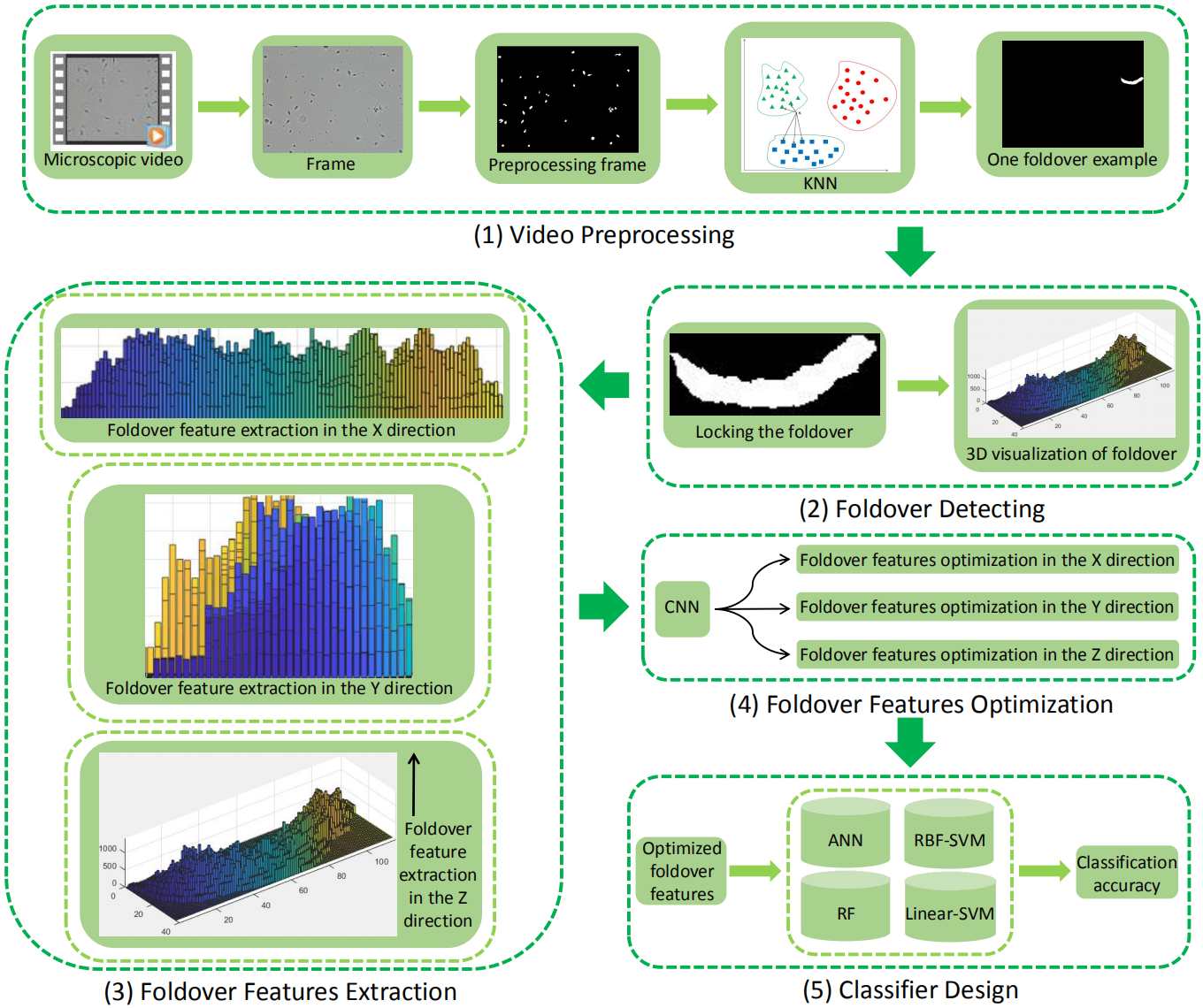}
\caption{The overview of proposed method mentioned in~\cite{li-2020-FFFD}. 
The figure corresponds to Fig.1 in the original paper.}
\label{fig:FF2}
\end{figure}

At present, the feature extraction of sperm micro video mainly adopts static features and 
dynamic features. However, due to the similarity of sperm cells, these two characteristics 
cannot be applied well. Our research team adopts a new feature extraction method, foldover feature \cite{li-2020-FFFD}, a behaviour description that helps analyse objects with weak visual information. The foldover feature is of great significance in analysing dynamic objects in microscopic videos. Foldover is defined as followning: Each frame of an object's motion is superimposed on the same spatial plane in the space-time order of the motion, the result of the superposition is the foldover of the object's motion. Foldover of an object contains temporal information, spatial information, behaviour features and static features. The flow chart of the foldover feature algorithm is presented in Fig.\ref{fig:FF2}.
\par

\begin{figure}[htbp!]
\centering
\includegraphics[width=0.35\linewidth]{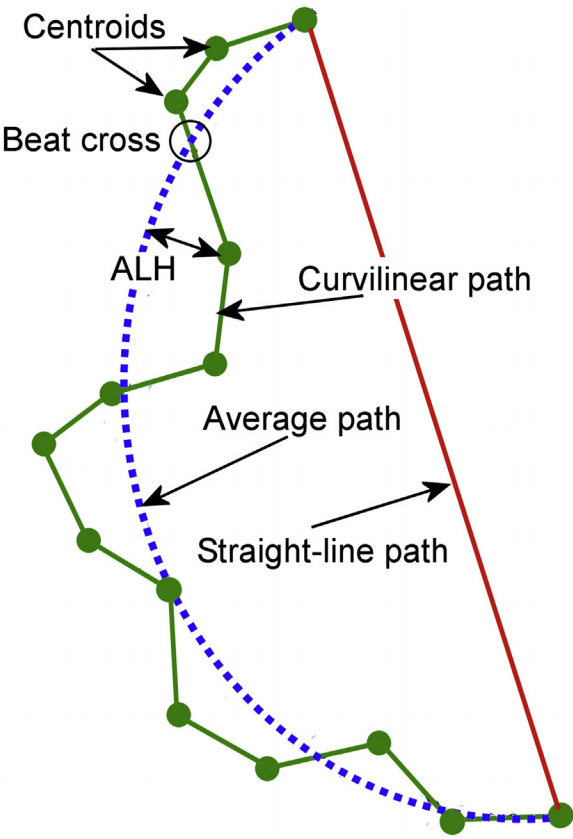}
\caption{Terminologies in CASA mentioned in \cite{amann-2014-CSAC}. The figure corresponds to Fig.2 in the original paper.} 
\label{fig:CS}
\end{figure}

Although the evaluation criteria of the CASA system are not unified, from the perspective 
of sperm movement, the movement of each sperm is a transfer of the position of the center 
of mass. And there is a uniform definition of the value of movement. In~\cite{amann-2014-CSAC}, some measured value are listed, seen in Fig.~\ref{fig:CS}. For example, curvilinear velocity [VCL], average path velocity [VAP], 
straight line velocity [VSL], amplitude of lateral head displacement [ALH], linearity of 
the curvilinear path [LIN], straightness of the average path [STR], and beat-cross frequency 
[BCF]. These values are essential for assessing sperm motility and can be used to rank levels of 
sperm health. Then, the percentage of motile sperm can be obtained. However, these motion values are rarely mentioned in the 
references above, and researchers focus more on whether the algorithm is accurate enough 
for sperm tracking. This may also be one of the limitations that prevent the CASA system 
from being used primarily for human reproductive health diagnosis.

\subsection{Potential Methods for CASA}
\label{ss:future:Potential}
Our review shows that the algorithms and technologies used by CASA are mostly classical 
methods, and not much application of the emerging algorithms. Therefore, we consulted 
some relevant technical literature of new algorithms and summarized some new methods that 
might be applied in the field of the CASA. In this way, the performance indexes of CASA 
system can be improved and the future development of CASA can be helped. 
\par
\paragraph{\textbf{Image Segmentation}}
In \cite{galvao-2020-ISUD}, a novel segmentation method combining sparse and dense method is proposed. Firstly, a sparse algorithm is applied for segmenting image, which can achieve faster segmentation than the super-pixel method. After that, a novel dense algorithm is applied for post-processing. By combining this two method, the proposed method achieve a good performance in image segmentation. Three natural image datasets are prepared for testing its performance. Results suggest that the proposed method yields  the same performance as the super-pixel method in a shorter time.
\par
In \cite{li-2020-SUIT},  a improved \textit{Unsupervised Image Translation} (SUIT) method is described for semantic segmentation.
 They adopt adversarial training for superior image generation, and design a novel semantic-content 
loss to enhance visual appearance preservation. Extensive experimental results demonstrate 
that these translated images within SUIT can significantly improve performance of the model 
on the target domain. Eventually this model with FCN8s-VGG16 architecture achieves around 
13 percentage points improvement in terms of mIoU on multiple semantic segmentation adaptation benchmarks.
\par
In \cite{han-2020-ACMI}, a novel active contour by employing Jeffreys divergence is proposed to deal with inhomogenous image segmentation. First, a improved local data fitting energy by applying Jeffreys divergence is presented for yielding a better performance on inhomogenous image segmentation. In addition, a novel global data fitting energy by applying Jeffreys divergence is presented for making versatility of proposed method higher. At last, the adaptive weights for two data fitting energies is designed.
\par
In \cite{liu-2020-ASAU}, an automatic segmentation method consisting of a neural network and image saliency is presented for 4D heart images. Experiment results suggest that the proposed method is able to achieve a good performance on 4D heart image segmentation within relatively short time.
\par
In \cite{pinheiro-2015-ANLP},  a novel segmentation method is presented for bone segmentation in medical image with high accuracy. The presented method mainly consists of two steps. The first one is pre-processing to obtain a rough segmentation image. The second step is further segmenting for image obtained in step one. In this step, deconvolution, cropping and interpolation operation are employed.
\par
In \cite{zhang-2020-AMCF,Zhang-2021-LANL}, a multiscale CNN-CRF (MSCC) system is described for environmental microorganisms (EMs) segmentation. MSCC model mainly consists of two parts. The first one is pixel-level segmentation part by employing mU-Net-B3. The second one is patch-level segmentation part by employing VGG-16. Experiment results on 420 EM samples suggest that by combining this two parts, the performance of EMs segmentation is greatly improved.
\par
In \cite{zhou-2020-AGIS}, an automatic segmentation method by applying affinity propagation (AP) clustering method is proposed for grayscale images. With the assistance of AP clustering method, image can be segmented without determining initial clustering center. In addition, AP clustering method can greatly improve the stability. Similarly, the clustering method in \cite{Li-2021-ACHI} is also possible for the sperm image segmentation.
\par
\begin{figure}[htbp!]
\centering
\includegraphics[width=0.75\linewidth]{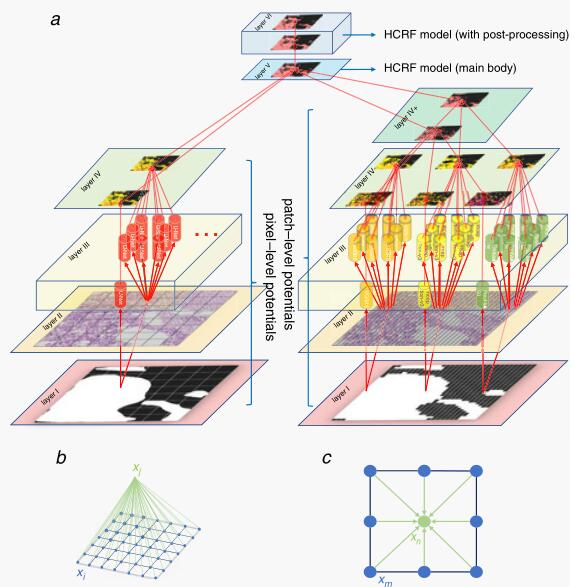}
\caption{The overview of proposed method mentioned in \cite{sun-2020-HCRF}. The figure corresponds to Fig.2 in the original paper.} 
\label{fig:1}
\end{figure}

In \cite{sun-2020-HCRF,Sun-2020-GHIS}, a new model named hierarchical conditional random field (HCRF) is described for segmenting gastric histopathology image. The overview of HCRF is shown in Fig.\ref{fig:1}. For pixel level potentials, a U-Net is retrained. For patch level potentials, small adjustments are made to VGG-16, Inception-V3, and ResNet-50. In addition, different weights are designed for different potentials for achieving a better performance. Experiment results on the validation set suggest that HCRF yields a high segmentation accuracy of 79.83$\%$.
\par

\paragraph{\textbf{Target Detection}}
In \cite{qiong-2020-ODBM}, the latest developments in target detection are summarized. In addition, the latest corresponding approaches in the field are tested and compared. In \cite{qiong-2020-ODBM}, the following contents are described: 1) introduction of latest related approaches; 2) listing the benchmark datasets and commonly used evaluation metrics; 3) experimental comparison of the performances.
\par
In \cite{raghunandan-2018-ODAV}, many target detection algorithms are Implemented and compared. The detection methods involved include those applied to various fields, such as shape detection, face detection. By comparing the detection performances of various detection algorithms, the most appropriate detection method is selected.
\par
In \cite{tripathi-2013-FAOB}, a method to detect abandoned objects from surveillance video is proposed. In the first step, foreground objects are extracted using background subtraction in which background modelling is done through running the average algorithm. After that, by analysing contour features of foreground objects in consecutive frames, static objects are detected. In the third step, detected static objects are classified into object and non-object objects using edge-based object recognition method.
\par
In \cite{mukherjee-2020-SBAT}, an automatic method is proposed for pulmonary nodules detection. By applying multi-level threshold, grayscale morphology and rolling ball algorithm, the features are extracted with higher accuracy. In addition, random under sampling method is employed for solving the problem of imbalance between classes.At last, binary particle swarm optimization is applied for increasing accuracy of detection result.
\par
\paragraph{\textbf{Object Tracking}}
In \cite{zhou-2020-DDLO}, a multiple object detection method is proposed. Based on a distractor-aware discrimination model, the proposed method is able to solve the problem of missed target detection. In addition, a relational attention learning mechanism is employed for the problem of object appearance variations. The validity of the proposed method is verified on the prepared data sets.

In~\cite{teng-2020-TASN}, by treating the tracking problem as a three-step decision-making 
process, a novel tracking network, which explores only three small subsets of candidate 
regions, is developed to achieve faster (real-time) localization of the target object along 
the frames in a video.

In \cite{madrigal-2020-3MTS}, a method for performing 3D motion tracking of the shoulder joint with respect to the thorax, using MARG sensors and a data fusion algorithm, is presented. 
\par
In \cite{zhang-2016-ATNS}, a novel system for tracking neural stem cells is proposed. Firstly, objects in image sequence are detected and localized. After that, the step of feature extraction is performed for the subsequent object tracking. At last, an improved mean shift method is proposed for statics objects.

In~\cite{mondal-2019-NMOT}, occlusion is one of the major challenges for object tracking 
in real-life scenarios. An effective observation model is proposed based on the confidence 
(classification) score provided to optimize the particle filter algorithm. 
\par
In \cite{wang-2020-MNMF}, an on-line object tracking algorithm based on he particle filtering framework is proposed. The object appearance is first modelled by subspace learning to reflect the target variations across frames. Furthermore, the alternating direction method of multipliers algorithm is employed to compute the model efficiently.

\subsection{Summary}
\label{ss:future:summary}
With advances in computer and Internet technology, the CASA system has great potential as 
a research tool in reproductive toxicology, animal production and human clinical analysis. 
Clearly, CASA needs to be rigorously tested to meet the standards of biological research 
and human reproductive medicine. In addition, CASA should pay more attention to sperm 
quality tests. Nowadays, male reproductive health diagnosis is highly subjective, so 
objective quantitative criteria are obviously urgently needed. Because this is more 
conducive to the quantification of sperm health analysis. It is not just looking at the 
motion tracking of sperm cells. CASA technology imperatively needs to improve the analysis 
of sperm motility, and incorporate it into the relevant experimental research. At the same 
time, the establishment of a large database with data analysis capability is also of great 
significance for the formulation of the diagnostic gold standard. These will improve the 
clinical relevance of semen analysis. More broadly, a stable and reliable CASA system will 
have a huge positive impact on natural life, wildlife conservation and animal husbandry.

\section{Conclusion and Future Work}
\label{s:Conclusion} 

Through the article, we can find that although there are various tracking algorithms, 
sperm detection method is still the most basic computer image processing method. Deep 
learning and neural networks are so prevalent and developed nowadays, but few research 
teams have applied them to the detection of sperm cells in the CASA system. Image 
recognition under the framework of deep learning has achieved high accuracy at present, 
and it is believed that it will also achieve good results when applied to CASA. Excellent 
cell detection results are an important prerequisite for sperm tracking algorithms. 
Therefore, it is possible to combine CASA technology with deep neural networks. In addition, 
most of the algorithms are only discussed at the research level in the laboratory, without 
practical application and clinical research. The main reason is also because of the lack 
of in-depth study and accumulation of sperm movement data. In the future, it will be 
important for the team to build a database, not just a sperm tracking algorithm. This will 
enable the CASA system to be applied to clinical medicine as soon as possible to reduce 
the subjectivity of diagnosis and the burden on doctors.
\section*{Acknowledgements}
This work was supported by the ``National Natural Science Foundation of China'' (No. 61806047), and the ``Fundamental Research Funds for the Central Universities'' (No. N2019003). They would also like to thank Miss Zixian Li and Mr. Guoxian Li for their important discussion.

\bibliographystyle{splncs04}      
\bibliography{Zhao}   

\end{document}